\documentclass[preprint,12pt,a4paper,twoside,fleqn,sort&compress]{elsarticle}
\usepackage{amssymb,latexsym,amscd,amsmath,amsthm}
\usepackage{pst-all}
\usepackage[varg]{pxfonts}
\usepackage{mathrsfs}
\usepackage{shuffle}
\usepackage{hyperref}
\usepackage{times}

\usepackage{color} 

\usepackage{mathrsfs}
\usepackage{graphicx}

\usepackage{helvet}
\usepackage{courier}

\usepackage{algorithm}
\usepackage[noend]{algorithmic}
\usepackage{verbatim}
\usepackage{flafter}
\newtheorem{definition}{Definition}
\newtheorem{theorem}{Theorem}

\newtheorem{remark}[theorem]{Remark}

\newtheorem{example}[theorem]{Example}
\newtheorem{proposition}[theorem]{Proposition}

\newcommand{\rw}{\rightarrow}

\newcommand{\sa}{\Sigma^{\ast}}

\begin{document}

\journal{arXiv}
\date{}

\begin{frontmatter}

\title{Model Checking of Linear-Time Properties in
Multi-Valued Systems
\thanks{This work is supported by National Science Foundation of China (Grant No: 11271237,61228305) and the Higher School Doctoral Subject Foundation of Ministry of Education of China (Grant No:200807180005).}}

\author{$^a$Yongming Li\corref{cor1}}
\ead{liyongm@snnu.edu.cn}
\cortext[cor1]{Corresponding Author}
\author{$^b$Manfred Droste}
\ead{droste@informatik.uni-leipzig.de}
\author{$^a$Lihui Lei}

\address{$^a$College of  Computer Science, Shaanxi Normal University, Xi'an, 710062, China}
\address{$^b$Institute of Computer Science, Leipzig University, D-04109 Leipzig, Germany}

\begin{abstract}
In this paper, we study the model-checking problem of linear-time properties in
multi-valued systems. Safety properties, invariant properties, liveness
properties, persistence and dual-persistence properties in
multi-valued logic systems are introduced. Some algorithms related
to the above multi-valued linear-time properties are discussed. The
verification of multi-valued regular safety properties and
multi-valued $\omega$-regular properties using lattice-valued
automata are thoroughly studied. Since the law of non-contradiction (i.e., $a\wedge \neg a=0$) and the law of excluded-middle (i.e., $a\vee \neg a=1$) do not hold in multi-valued logic, the linear-time properties introduced in this paper have new forms compared to those in classical logic.
Compared to those classical model-checking methods, our methods to multi-valued model checking are  accordingly more
direct: We give an algorithm for showing $TS\models P$ for a model $TS$ and a linear-time property $P$, which proceeds by directly checking the inclusion $Traces(TS)\subseteq P$ instead of $Traces(TS)\cap \neg P=\emptyset$. A new form of multi-valued model checking with membership
degree is also introduced. In particular, we show that multi-valued
model checking can be reduced to classical model checking. The
related verification algorithms are also presented. Some illustrative examples and a case study are also provided.
\end{abstract}

\end{frontmatter}

\section{Introduction}
\label{intro}

In the last four decades, computer scientists have systematically
developed theories of correctness and safety as well as
methodologies, techniques and even automatic tools for correctness
and safety verification of computer systems; see for example
\cite{lamport77,manna95,alpern85}. Of which, model checking has
become established as one of the most effective automated techniques
for analyzing correctness of software and hardware designs. A model
checker checks a finite-state system against a correctness property
expressed in a propositional temporal logic such as LTL (Linear Temporal Logic) or CTL (Computational Tree Logic).
These logics can express safety (e.g., ``No two processes can be in
the critical section at the same time'') and liveness (e.g., ``Every
job sent to the printer will eventually print'') properties.
Model checking has been effectively applied to reasoning about
correctness of hardware, communication protocols, software
requirements, etc. Many industrial model checkers have been
developed, including SPIN \cite{holzmann97}, SMV \cite{mcmillan93}.

Despite their variety, existing model checkers are typically limited
to reasoning in classical logic. However, there are a number of
problems for which classical logic is insufficient. One of these is
reasoning under uncertainty. This can occur either when complete
information is not known or cannot be obtained (e.g., during
`requirements' analysis), or when this information has been removed
(abstraction). Classical model checkers typically deal with
uncertainty by creating extra states, one for each value of the
unknown variable and each feasible combination of values of known
variables. However, this approach adds significant extra complexity
to the analysis. Classical reasoning is also insufficient for models
that contain inconsistencies. Models may be inconsistent because they
combine conflicting points of view, or because they contain
components developed by different people. Conventional reasoning
systems cannot cope with inconsistency because the presence of a
single contradiction results in trivialization -- anything follows
from $A\wedge \neg A$. Hence, faced with an inconsistent description
and the need to perform automated reasoning, we must either discard
information until consistency is achieved again, or adopt a
nonclassical logic. Multi-valued logic (mv-logic, in short) provides
a solution to both reasoning under uncertainty and under
inconsistency. For example, we can use ¡°unknown¡± and ¡°no
agreement¡± as logic values. In fact, model checkers based on
three-valued and four-valued logics have already been studied. For
example, \cite{chechik00} (c.f., \cite{shoham10}) used a three-valued logic for interpreting
results of model-checking with abstract interpretation, whereas
\cite{hazelhurst96} used four-valued logics for reasoning about
abstractions of detailed gate or switch-level designs of circuits.
For reasoning about dynamic properties of systems, we need to extend
existing modal logics to the multi-valued case. Fitting
\cite{fitting91} explores two different approaches for doing this:
the first extends the interpretation of atomic formulae in each
world to be multi-valued; the second also allows multi-valued
accessibility relations between worlds. The latter approach is more
general, and can readily be applied to the temporal logics used in
model checking \cite{chechik012}. We use different multi-valued
logics to support different types of analysis. For example, to model
information from multiple sources, we may wish to keep track of the
origin of each piece of information, or just the majority vote, etc.
Thus, rather than restricting ourselves to any particular
multi-valued logic, our approach is to extend classical symbolic
model checking to arbitrary multi-valued logics, as long as
conjunction, disjunction and negation of the logical values are well
defined. M. Chechik and her colleagues have published a series of papers along this line, see \cite{chechik00,chechik01,chechik04,chechik012,chechik06}.

Our purpose is to develop automata-based model-checking techniques
in the multi-valued setting. More precisely, the major design decision
of this paper is as follows:

A lattice-valued automaton is adopted as the model of the systems.
This is reasonable since classical automata (or equivalent
transition systems) are common system models in classical
model checking. Linear-time properties of multi-valued systems
are checked in this paper. They are defined to be infinite sequences of
sets of atomic propositions, as in the classical case, with
truth-values in a given lattice. The key idea of the automata-based
approach to model checking is that we can use an auxiliary automaton
to recognize the properties to be checked, and then combine it
with the system to be checked so that the problem of checking the
safety or $\omega$-properties of the system is reduced to checking
some simpler (invariance or persistence) properties of the larger
system composed by the systems under checking and the auxiliary
automaton. A difference between the classical case and the
multi-valued case deserves a careful explanation. Since the law
of non-contradiction (i.e., $a\wedge \neg a=0$) and the law of
excluded middle (i.e., $a\vee \neg a=1$) do not hold in multi-valued
logic, the present forms of many classical properties in
multi-valued logic must have some new forms, and some distinct
constructions need to be given in multi-valued logic.

As said in Ref. \cite{baier08}, the equivalences and
preorders between transition systems that ``correspond'' to
linear temporal logic are based on trace inclusion and equality,
whereas for branching temporal logic such relations are based on
simulation and bisimulation relations. That is to say, the model checking of
a transition system $TS$ which represents the model of a system
satisfying a linear temporal formula $\varphi$, i.e., $TS\models \varphi$
is equivalent to checking the inclusion relation
$Traces(TS)\subseteq P$, where $Traces(TS)$ is the trace function
of the transition system $TS$ and $P$ is the temporal property
representing the formula $\varphi$. In classical logic, we know
that $a\leq b$ if and only if $a\wedge \neg b=0$ holds. Therefore,
$TS\models \varphi$ if and only if $Traces(TS)\cap \neg P=\emptyset$.
Then, instead of checking $TS\models \varphi$ directly using the inclusion relation $Traces(TS)\subseteq P$,
it is equivalent to checking the
emptiness of the language $L({\cal A})\cap L({\cal
A}_{\neg\varphi})$ indirectly, where ${\cal A}$ is a B\"{u}chi automaton
representing the trace function of the transition system $TS$
(i.e., $L({\cal A})=Traces(TS))$, and ${\cal A}_{\neg\varphi}$ is
a B\"{u}chi automaton related to temporal property $\neg\varphi$
(i.e., $L({\cal A}_{\neg\varphi})=\neg P$).

In contrast, in
mv-logic, $a\leq b$ is in general not equivalent to the condition
$a\wedge \neg b=0$, so the classical method to solve model
checking of linear-time properties does not universally apply to the multi-valued model checking.
The available methods of multi-valued model checking
(\cite{chechik01}) still used the classical method with some minor
correction. That is, instead of checking of $TS\models P$ for a
multi-valued linear time property $P$ using the inclusion of the trace
function $Traces(TS)\subseteq P$, the available method only
checked the membership degree of the language $L({\cal A})\cap
L({\cal A}_{\neg P})$, where ${\cal A}_{\neg P}$ is a
multi-valued B\"{u}chi automaton such that $L({\cal A}_{\neg P})=\neg P$. As we know, these two methods are not equivalent in mv-logic.
Then, some new methods to apply multi-valued model checking of linear-time
properties based on trace inclusion relations need to be
developed.

We provide new results along this line. In
fact, we shall give a method of multi-valued model checking of linear-time
property directly using the inclusion of the trace function of
$TS$ into a linear-time property $P$. In propositional logic, we know
that we can use the implication connective $\rw$ to represent the
inclusion relation. In fact, in classical logic, we know that the
implication connective can be represented by disjunction and
negation connectives, that is, $a\rw b=\neg a\vee b$. In this
case, we know that $a\leq b$ if and only if $\neg a\vee b=1$, if
and only if $a\wedge \neg b=0$, if and only if $a\rw b=1$. Then a natural problem arises: how to
define the implication connective in multi-valued logic? By the above analysis,
it is not appropriate to use the implication connective
defined in the form $a\rw b=\neg a\vee b$ to represent the
inclusion relation in multi-valued logic. In order to use the implication connective to reflect the inclusion relation in mv-logic, we
shall use implication connective $\rw$ as a primitive connective in
multi-valued logic as done in \cite{hajek97}. In this case, we
will have that $a\leq b$ is equivalent to $a\rw b=1$ semantically. Then
we can use the implication connective to present the inclusion relation
in multi-valued logic. This view will give a new idea to study
linear-time properties in multi-valued model checking.
Furthermore, we also show that we can use the classical model
checking methods (such as SPIN and SMV) to solve the multi-valued model-checking problem.
In particular, some special and important
multi-valued linear-time properties are introduced, which include
safety, invariance, persistence and dual-persistence properties,
and the related verification algorithms are also presented. In
multi-valued systems, the verification of the mentioned properties
require some different structures compared to their classical counterpart. In
particular, since the law of non-contradiction and the law of
excluded middle do not hold in multi-valued logic, the auxiliary
automata used in the verification of multi-valued regular safety
properties and multi-valued $\omega$-regular properties need to be
deterministic, whereas nondeterministic automata suffice for
the classical cases.

There are at least two  advantages of the
method used in this paper. First, we use the implication
connective as a primitive connective which can reflect the ``trace
inclusion'' in multi-valued logic, i.e., in multi-valued model checking, $TS\models P$ if and only if $Trace(TS)\subseteq P$, the natural corresponding counterpart in multi-valued logic is,  $a\leq b$ if and only if
$a\rw b=1$. Second, since there is a well-established multi-valued
logic frame using the implication connective as a primitive connective
(\cite{hajek97}), there will be a nice theory of multi-valued
model checking, especially, model checking of linear-time property in mv-logic.
Of course, this approach can be seen as another view on the
study of multi-valued model checking.

The content of this paper is arranged as follows. We first recall
some notions and notations in multi-valued logic systems in
Section 2. In Section 3, the multi-valued linear-time properties
are introduced. In particular, the notions of multi-valued regular safety properties and
multi-valued liveness properties are introduced, then the reduction of
model checking of multi-valued invariant properties into classical ones is
presented. The verification of multi-valued regular safety properties is
discussed in Section 4. In Section 5, the verification of
multi-valued $\omega$-regular properties is developed. Some general
considerations about the multi-valued model checking are discussed
in Section 6, in which the truth-valued degree of an mv-transition
system satisfying a multi-valued linear-time property is
introduced. Examples and a case study illustrating the method of this article are
presented in Section 7. The summary, comparisons and the future work are included in the
conclusion part. We place the proofs of some propositions of this
article in the Appendix parts for readability.

\section {Multi-valued logic: some preliminaries}
\label{sec:1}

Let us first recall some notions and notations of multi-valued
logic, which can be found in the literature
\cite{birkhoff40,chechik04,bloc92,hajek97}.
We start by presenting ordered sets and lattices which play a very
important role in multi-valued logic.

\begin{definition}\label{de:poset}

{\rm A {\sl partial order}, $\leq$, on a set $l$ is a binary relation on
$l$ such that for all $x, y, z\in l$ the fo1lowing conditions hold:

(1)\ (reflexivity) $x\leq x$.

(2)\ (anti-symmetry) $x\leq y$ and $y\leq x$ imply $x=y$.

(3)\ (transitivity) $x\leq y$ and $y\leq z$ imply $x\leq z$.}

\end{definition}

A partially ordered set, $(l,\leq)$, has a {\sl bottom} (or {\sl the least})
element if there exists $0\in l$ such that $0\leq x$ for any $x\in
l$. The bottom element is also denoted by $\bot$. Dually, $(l,\leq)$
has a {\sl top} (or {\sl the largest}) element if there exists $1\in l$ such
that $x\leq 1$ for all $x\in l$. The top element is also denoted as
$\top$.

\begin{definition}\label{de:lattice}

{\rm A partially ordered set, $(l,\leq)$, is a {\sl lattice} if the
greatest lower bound and the least upper bound exist for any nonempty finite
subset of $l$.}

\end{definition}

Given lattice elements $a$ and $b$, their greatest lower bound is
referred to as {\sl meet} and denoted $a\wedge b$, and their least upper
bound is referred to as {\sl join} and denoted $a\vee b$. By Definition
\ref{de:lattice}, a lattice $(l,\leq)$ is called bounded if it contains a top element $1$ and a
bottom element $0$.

\begin{remark}\label{re:complete lattice}

{\rm A {\sl complete lattice} is a partially ordered set, $(l,\leq)$, in
which the greatest lower bound and the least upper bound exist for
any  subset of $l$. For a subset $X$ of $l$, its greatest lower
bound and least upper bound are denoted by $\bigwedge X$ or $\bigvee
X$, respectively. Any complete lattice is bounded, since $1=\bigwedge\emptyset$ and $0=\bigvee\emptyset$.}

\end{remark}

\begin{definition}\label{de:distributive lattice}

{\rm A lattice $l$ is {\sl distributive} if and only if
one of the following (equivalent) distributivity laws holds,

$x\wedge (y\vee z)=(x\wedge y)\vee (x\wedge z)$,

$x\vee (y\wedge z)=(x\vee y)\wedge (x\vee z)$.}

\end{definition}

The join-irreducible elements are crucial for the use of
distributive lattices in this article.

\begin{definition}\label{de:join-irreducible}

{\rm Let $l$ be a lattice. Then an element $x\in l$ is called {\sl join-irreducible}
if $x\not=0$ and $x=y\vee z$ implies $x=y$ or $x=z$ for all $y,z\in
l$.}

\end{definition}

If $l$ is a distributive lattice, then a non-zero element $x$ in
$l$ is join-irreducible iff  $x\leq y\vee z$ implies that $x\leq
y$ or $x\leq z$ for any $y,z\in l$. We use $JI(l)$ to denote the
set of join-irreducible elements in $l$. It is well-known
that $l$ is generated by its join-irreducible elements if $l$ is a finite
distributive lattice, that is, for any $a\in l$, there exists a finite
subset $A$ of $JI(l)$ such that $a=\bigvee A$. In other words,
every element of $l$ can be written as the join of finitely many
join-irreducible elements.

Furthermore, we present the definition of de Morgan algebra, also
called quasi-Boolean algebra as in \cite{chechik04}.

\begin{definition}\label{de:de Morgan algebra}

{\rm A {\sl de Morgan algebra} is a tuple $(l, \leq, \wedge, \vee, \neg,
0, 1)$, such that $(l, \leq, \wedge, \vee, 0,1)$ is a bounded distributive
lattice, and the negation $\neg$ is a function $l\rw l$ such that
$x\leq y$ implies $\neg y\leq \neg x$ and $\neg\neg x=x$ for any
$x,y\in l$. Then $\neg x$ is also called the (quasi-)complement of $x$.}

\end{definition}

In a de Morgan algebra, the de Morgan laws hold, that is,
$\neg(x\vee y)=\neg x \wedge \neg y$ and $\neg(x\wedge y)=\neg x\vee
\neg y$. Also, $\neg 0=1$ and $\neg 1=0$. It is well-known that a Boolean algebra is a de Morgan algebra $B$
with the additional conditions that for every element $x\in B$,

Law of Non-Contradiction:  \ \ $x\wedge \neg x=0$.

Law of Excluded Middle: \ \ \ \ \ $x\vee\neg x=1$.

\begin{example}\label{ex:algebras}

{\rm In Fig. 1, we present some examples of de Morgan algebras, where
$B_2$, $l_3$ and $l_5$ are linear orders.

(1) The lattice $B_2$ in Fig.1, with $\neg 0=1$ and $\neg 1=0$, gives us classical logic.

(2) The three-valued logic $l_3$ is defined in Fig.1, where $\neg$F=T, $\neg$M=M and $\neg$T=F.

(3) The lattice $B_2\times B_2$ in Fig.1 shows the product algebra, where $\neg(0,0)=(1,1)$, $\neg(1,0)=(0,1)$, $\neg(0,1)=(1,0)$ and
$\neg(1,1)=(0,0)$. This logic can be used for reasoning about disagreement between two knowledge sources.

(4) The lattice $l_5$ in Fig.1 shows a five-valued logic and possible interpretations of its value as, T=Definitely true,
L=Likely or weakly true, M=Maybe or unknown, U=Unlikely or weakly false, and F=Definitely false,
where $\neg$T=F, $\neg$L=U, $\neg$M=M, $\neg$U=L, and $\neg$F=T.

(5) The lattice $l_3\times l_3$ in Fig.1 shows a nine-valued logic constructed as the
product algebra. Like $B_2\times B_2$, this logic can be used for reasoning about
disagreements between two sources, but also allows missing information
in each source.}

\end{example}

\begin{figure}[ptb]
\begin{center}
\includegraphics[width=\textwidth]{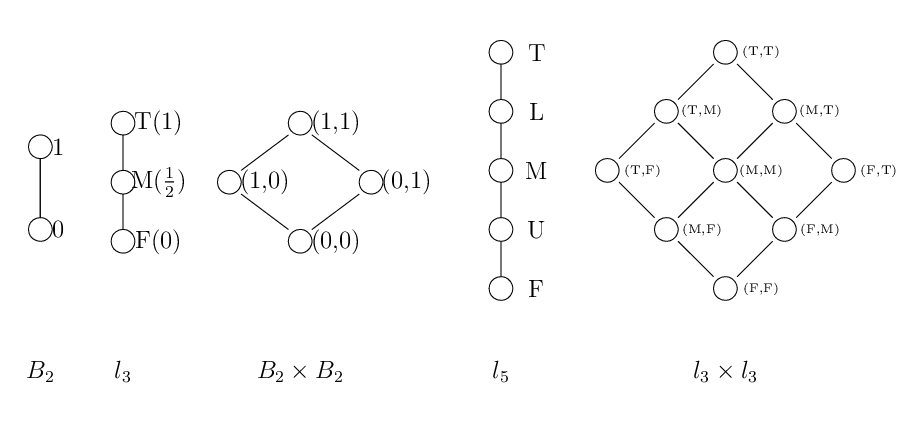}
\end{center}
\caption{Some lattices}
 \label{fig:mc1}
 \end{figure}


In the following, we always assume that $l$ is a de Morgan algebra,
and it is also called an algebra.

Given an algebra $l$, we now can define multi-valued sets and multi-valued relations,
which are functions taking values in $l$.
Multi-valued sets and multi-valued relations are basic data structures in multi-valued model checking introduced later in this paper.

\begin{definition}\label{de:mv-set}

{\rm Given an algebra $l$ and a classical set $X$, an {\sl $l$-valued set}
on $X$, referred to as $f$, is a function $X\rw l$.

The collection of all $l$-sets on $X$ is denoted $l^X$, called the $l$-power set of $X$.}

\end{definition}

When the underlying algebra $l$ is clear from the context, we refer to
an $l$-valued set just as {\sl multi-valued set} (mv-set, for short).
For an mv-set $f$ and an element $x$ in $X$, we will use $f(x)$ to
define the membership degree of $x$ in $f$. In the classical case,
this amounts to representing a set by its characteristic function.

The standard operations on mv-sets $f, g$ are defined in the following manner:

{\sl mv-intersection}: $(f \cap g)(x)\triangleq f(x)\wedge g(x)$.

{\sl mv-union}: $(f \cup g)(x)\triangleq f(x)\vee g(x)$.

{\sl set inclusion}: $f\subseteq g \triangleq\forall x.(f(x)\leq g(x))$.

{\sl extensional equality}: $f= g \triangleq\forall x.(f(x)= g(x))$.

{\sl mv-complement}: $\neg f(x) \triangleq \neg (f(x))$.

\begin{definition}\label{de:mv-relation}

{\rm For a given algebra $l$, an {\sl $l$-valued relation} $R$ on two sets
$X$ and $Y$ is an $l$-valued set on $X\times Y$.}

\end{definition}

For any $l$-valued set $f:X\rw l$, and for any $m\in l$, the {\sl
$m$-cut} of $f$ is defined as the subset $f_m$ of $X$ with

$f_m=\{x\in X | f(x)\geq m\}$.

\noindent The {\sl support} of $f$, denoted by $supp(f)$, is the following subset
of $X$,

$supp(f)=\{x\in X | f(x)>0\}$.

Then we have a resolution of $f$ by its cuts presented in the
following proposition.

\begin{proposition}\label{pro:resolution}

For any $l$-valued set $f:X\rw l$, we have

$f=\bigcup_{m\in l} m\wedge f_m$,

\noindent where $m\wedge f_m$ is an $l$-valued set defined as
$m\wedge f_m(x)=m$ if $x\in f_m$ and $0$ otherwise.
Furthermore, if $l$ is finite, then

$f=\bigcup_{m\in JI(l)} m\wedge f_m$.

\end{proposition}

The verification is simple, we omit its proof here. As a corollary,
we have the following proposition.

\begin{proposition}\label{pro:level of leq}

Given two $l$-valued sets $f,g: X\rw l$, $f\leq g$ if and only if
$f_m\subseteq g_m$ for every $m\in l$. Furthermore, if $l$ is
finite, $f\leq g$ if and only if $f_m\subseteq g_m$ for every
$m\in JI(l)$.

\end{proposition}

In order to define the semantics of multi-valued implication, we will need the algebra $l$ to have an implication operator. There are at least two methods to define the implication operator. First, it can be defined by other primitive connectives in mv-logic logic. For example, we can use $a\rw b=\neg a\vee b$ as a material implication or $a\rw b=\neg a\vee (a\wedge b)$ as a quantum logic implication to define the implication operator. In fact, in
Ref.\cite{chechik04,chechik01}, the implication operator is chosen
as the material implication. The second choice of implication operator is
as a primitive connective in $l$ that satisfies
the condition $a\rw b=1$ whenever $a\leq b$. In this paper, we shall use the second method to define the implication operator. We shall give some analysis of our choice in Section 6. Then we need $l$ to be a residual lattice or Heyting algebra defined as follows.

\begin{definition}\label{de:residula lattice}

{\rm Let $l$ be a bounded lattice and $\rw$ a binary function on $l$ such that for any $a,b\in l$, the element $a\rw b=\rw(a,b)$  in $l$ satisfies the following condition,

$x\leq a\rw b$ iff $x\wedge a\leq b$,

\noindent for any $x\in l$. Then $l$ is called a residual lattice or Heyting algebra, and the operator $\rw$ is called the implication or the residual operator in $l$.}

\end{definition}

For example, if $l$ is a linear order, then $a\rw b=1$ if $a\leq b$
and $a\rw b=b$ if $a>b$; if $l$ is a Boolean algebra, then $a\rw
b=\neg a\vee b$. In particular, each finite distributive lattice is a residual lattice. Note that in any residual lattice, we have $a\rw b=1$ iff $a\leq b$.

Any complete lattice $l$ satisfying the infinite distributive law, i.e.,

$x\wedge(\bigvee_{i\in I}a_i)=\bigvee_{i\in I}(x\wedge a_i)$,

\noindent is a residual lattice, and the implication operator is defined as follows,

$a\rw b=\bigvee\{c\in l | a\wedge c\leq b\}$.

The algebra $l$ in this paper is required to be a residual lattice. This is the main difference of our method from those used in \cite{chechik00,chechik01,chechik012,chechik04,chechik06}. We shall give some analysis why we use the implication operator in the second form in Section 6.

\begin{remark}\label{re: residual lattice}
{\rm As ordered structures we take Heyting algebras which are
de Morgan algebras, i.e., bounded lattices which have an
residual operator $\rw$  and a self-inverse negation operation.
It is known from lattice theory that there are many Heyting
and de Morgan algebras which are not Boolean algebras,
cf., e.g., \cite{gratzer03}. For instance, any finite linear order is
a Heyting and de Morgan algebra but not a Boolean algebra
(if it has more than 2 elements). Other examples of
Heyting algebras occur e.g. in intuitionistic logic and in
pointless topology studied for denotational semantics of
programming languages.}
\end{remark}

With these preliminaries, we can introduce some simple facts about
multi-valued logic (mv-logic, in short).

Similar to that of classical first-order logic, the syntax of
multi-valued or $l$-valued logic has three primitive connectives
$\vee$ (disjunction), $\neg$ (negation) and $\rw$ (implication),
and one primitive quantifier $\exists$ (existential quantifier).
In addition, we need to use some set-theoretical formulas. Let
$\in$ (membership) be a binary (primitive) predicate symbol. Then
$\subseteq$ and $\equiv$ (equality) can be defined with $\in$ as
usual. The semantics of multi-valued logic is given by
interpreting the connectives $\vee$ and $\neg$ as the operations
$\vee$ and $\neg$ on $l$, respectively, and interpreting the
quantifier $\exists$ as the least upper bound in $l$. Moreover,
the truth value of the set-theoretical formula $x\in A$ is $[x\in
A]=A(x)$. In multi-valued logic, $1$ is the unique designated
truth value; a formula $\varphi$ is valid iff $[\varphi]=1$, and
denoted by $\models_l\varphi$.


In this article, we only use multi-valued proposition formulae. We
give their formal definition here.

\begin{definition}\label{de:mv-proposition}

{\rm Given a set of atomic propositions $AP$, the {\sl multi-valued
proposition formulae} ({\sl mv-proposition formulas}, in short) generated
by $AP$ are defined by the following BNF expression:

$\varphi:= A|r|\varphi_1\vee \varphi_2|\neg\varphi |\varphi_1\rw
\varphi_2,$

\noindent where $r\in l$ and $A\in AP$.

The set of mv-proposition formulae is denoted by $l$-$AP$.}

\end{definition}

We can define conjunction and equivalence as usual,

$\varphi_1\wedge\varphi_2=\neg(\neg\varphi_1\vee \neg\varphi_2)$ and
$\varphi_1\leftrightarrow \varphi_2=(\varphi_1\rw \varphi_2)\wedge
(\varphi_2\rw \varphi_1)$.


For any valuation of atomic propositions $v: AP\rw l$, the
truth-value of an mv-proposition formula $\varphi$ under $v$ is an
element in $l$, denoted $v(\varphi)$, which is defined inductively
as follows,

$v(\varphi)=v(A)$ if $\varphi=A\in AP$;

$v(\varphi)=r$ if $\varphi=r\in l$;

$v(\varphi_1\vee\varphi_2)=v(\varphi_1)\vee v(\varphi_2)$;

$v(\neg\varphi)=\neg v(\varphi)$;

$v(\varphi_1\rw\varphi_2)=v(\varphi_1)\rw v(\varphi_2)$.

For a set of proposition formulae $\Phi\subseteq AP$, the
characterization function of $\Phi$ is a valuation $v$ on $AP$ such
that $v(A)=1$ if $A\in \Phi$ and $0$ otherwise. In this case, we
write $v(\varphi)$ as $\varphi(\Phi)$.

Multi-valued temporal logic formulae have also been defined in the literature. For further reading, we refer to \cite{chechik04}.

\section {Linear-time properties in multi-valued systems}
\label{sec:2}

In this section, we shall introduce several notions of linear-time
properties in mv-logic, including multi-valued version of safety,
invariance, persistence, dual
persistence, and liveness. As starting point, let us first give the notion of
multi-valued transition system, which is used to model the system
under consideration.

\subsection{Multi-valued transition systems and their trace
functions}

Transition systems or Kripke structures are the key models for model
checking. Corresponding to multi-valued model checking, we
have the notion of multi-valued transition systems, which are defined as
follows (for the notion of multi-valued Kripke structures, we refer
to \cite{chechik04}).

\begin{definition}\label{def:mv-ts}
{\rm A {\sl multi-valued transition system} (mv-TS, for short) is a
6-tuple $TS=(S,Act,\eta, I,AP, L)$, where

(1) $S$ denotes a set of
states;

(2) $Act$ is a set of the names of actions;

(3)
$\eta: S\times Act\times S\rw l$ is an mv-transition relation;

(4) $I: S\rw l$ is mv-initial state;

(5) $AP$ is a set of
(classical) atomic propositions;

(6) $L: S\rw 2^{AP}$ is a
labeling function.

$TS$ is called finite if $S$, $Act$,and $AP$ are finite.

We always assume that an mv-TS is finite in this paper.}

\end{definition}

Here, the labeling function $L$ is the same as in the classical
case. In Ref.\cite{chechik04}, it required that the labeling
function is also multi-valued, that is, $L$ is a function from
the states set $S$ into $l^{AP}$. We shall show that
they are equivalent as trace functions in Appendix A.

For convenience, we use $(s,\alpha,s^{\prime},r)\in\rw$ to represent
$\eta(s,\alpha,s^{\prime})=r$, and the $TS=(S,Act,\eta, I,AP, L)$ is denoted by $TS=(S,Act,\rw, I,AP, L)$ in the following. Intuitively,
$\eta(s,\alpha,s^{\prime})$ stands for the truth value of the
proposition that action $\alpha$ causes the current state $s$ to
become the next state $s^{\prime}$. The intuitive behavior of an
mv-transition system can be described as follows. The transition
system starts in some initial state $s_0\in I$ (in multi-valued
logic) and evolves according to the transition relation $\rw$.
That is, if $s$ is the current state, then a transition
$(s,\alpha,s^{\prime},r)\in\rw$ originating from $s$ is selected
in the mv-logic sense and taken, i.e., the action $\alpha$ is
performed and the transition system evolves from state $s$ into
state $s^{\prime}$ with truth value $r$. This selection
procedure is repeated in state $s^{\prime}$ and finishes once a
state is encountered that has no outgoing transitions. (Note that
$I$ may be empty; in that case, the transition system has no
behavior at all as no initial state can be selected.) It is
important to realize that in case a state has more than one
outgoing transition, the ¡°next¡± transition is chosen in a purely
mv-logic fashion. That is, the outcome of this selection process
is known with some truth-value a priori, and, hence, the degree
with which a certain transition is selected is given a priori in the mv-logic sense.

Let $TS =(S,Act,\rw,I,AP,L)$ be a transition system. A finite
{\sl execution fragment} (or a {\sl run}) $\varrho$ of TS is an
alternating sequence $\varrho=s_0\alpha_1 s_1\alpha_2 ...\alpha_ns_n$ of states and actions ending with a state.
If $\eta(s_i,\alpha_{i+1},s_{i+1})=r_{i+1}$ for all $0\leq i<n$,
where $n \geq 0$, the sequence has truth value
$v(\varrho)=I(s_0)\wedge r_1\wedge r_2\wedge\cdots\wedge r_n$. We
refer to $n$ as the length of the execution fragment $\varrho$. An
infinite execution fragment $\rho$ of $TS$ is an infinite,
alternating sequence of states and actions:
$\rho=s_0\alpha_1 s_1\alpha_2 ...$, and if
$\eta(s_i,\alpha_{i+1},s_{i+1})=r_{i+1}$ for all $0\leq i$,
the sequence has truth value $v(\rho)=I(s_0)\wedge r_1\wedge
r_2\wedge\cdots=\bigwedge_{i\geq 0}r_i$, where $r_0=I(s_0)$.

For a finite execution fragment $\varrho$ or an infinite execution
fragment $\rho$ of $TS$, the corresponding finite sequence or
infinite sequence of states, denoted $\pi(\varrho)=s_0s_1\cdots s_n$
or $\pi(\rho)=s_0s_1\cdots$, respectively, is called the {\sl path} of $TS$
corresponding to $\varrho$ or $\rho$.

In general, an {\sl infinite path} or a {\sl computation} of an mv-TS, $TS$, is
an infinite sequence of states (i.e., $s_0s_1\cdots$) such that
$s_0\in I$ and $\eta(s_i,\alpha_i, s_{i+1})>0$ for some $\alpha_i$.
In order to describe an infinite sequence of states, we will use the
function $\pi: {\bf N}\rw S$ defined as: $\pi(i)$ is the i-th state
in the sequence $s_0s_1\cdots$. In the following, $\pi$ will denote
a path of the mv-TS and $\pi[i]$ will denote the actual sequence of
states, that is, $\pi[i]=\pi(i)\pi(i+1)\cdots$. We use
$\overline{\pi}$ to denote a finite fragment of $\pi$.

Let $TS=(S,Act,\rw,I,AP, L)$ be an mv-TS, then for each $s\in S$,

$Paths_{TS}(s)=\{\pi: {\bf N}\rw S | (\pi(0)=s)(\forall i\in {\bf
N})(\exists \alpha_i\in Act)(\eta(\pi(i),\alpha_i,\pi(i+1))> 0)\}$,

\noindent which is the set of all infinite paths starting at state
$s$.

For $T\subseteq S$, we write $Paths_{TS}(T)=\bigcup_{s\in
T}Paths_{TS}(s)$. Let $Paths(TS)=Paths_{TS}(S)$.

Also, we define $S_{inf}=\{ s\in S | Paths_{TS}(s)\not=\emptyset\}$.
If the transition relation $\rw$ is total, that is, for all $s\in
S$, there exists $\alpha\in Act$ and $s^{\prime}\in S$ such that
$\eta(s,\alpha,s^{\prime})>0$, then we also call this $TS$ {\sl
without terminal state}. In this case, $S_{inf}=S$.

A {\sl trace} is the sequence of labelings (or observations)
corresponding to a path $\pi$, $L(\pi(0))L(\pi(1))\cdots$ which
will be again denoted by $L(\pi)$ or $trace(\pi)$. The definition
of the trace as function will be the composition of the map $L$
and $\pi$, i.e., the map $L\circ \pi: {\bf N}\rw 2^{AP}$. The {\sl $l$-language}
or {\sl multi-valued language} (mv-language, in short) of the
transition system $TS$ over $2^{AP}$, which is also called the
{\sl multi-valued trace function} of $TS$, is defined as the
function $Traces(TS)$ from $(2^{AP})^{\omega}$ into $l$ as follows,

$Traces(TS)(\sigma)=\bigvee\{v(\rho) | L(\pi(\rho))=\sigma\}$.

\noindent Observe that this supremum exists since by assumption $TS$ is finite, hence $v$ has finite image. In fact, $Traces(TS)$ registers sequences of the set of
atomic propositions $L(\pi)$ that are valid along the execution
with truth value $Traces(TS)(L(\pi))$.

A multi-valued trace function $Traces(TS):
(2^{AP})^{\omega}\rw l$ is a multi-valued linear-time
property over $2^{AP}$ defined in general as follows.

\begin{definition}\label{de:LTproperty}

{\rm An {\sl mv-linear-time property} ({\sl LT-property}, in short) $P$ over the set of atomic
propositions $AP$ is an mv-subset of $(2^{AP})^{\omega}$, i.e., $P:
(2^{AP})^{\omega}\rw l$.}

\end{definition}

LT properties specify the traces that an mv-TS should exhibit.
Informally speaking, one could say that an LT property specifies
the admissible (or desired) behavior of the system under
consideration.

The fulfillment of an LT property by an mv-TS is defined as
follows.

\begin{definition}\label{de:sat}

{\rm For an mv-TS, $TS$, and an mv-linear-time property $P$,
we let $TS\models P$ if $Traces(TS)\subseteq P$.}

\end{definition}

In mv-logic, even if $TS\models P$ does not hold, i.e.,
$Traces(TS)\subseteq P$ does not hold, we still have the membership degree of
the inclusion relation, denoted $lMC(TS, P)$, which presents the
degree of the inclusion of $Traces(TS)$ in $P$. The study of $lMC(TS,
P)$ is more general and complex, so we will discuss it only in Section 6.

In the following, we will define several mv-linear-time properties
including safety and liveness properties.

\subsection{Multi-valued safety property}

Safety properties are often characterized as ``nothing bad should
happen''. Formally, in the classical case, a safety property is defined
as an LT property over $AP$ such that any infinite word $\sigma$
where $P$ does not hold contains a bad prefix. Since it is
difficult to define the notion of {\sl bad prefix} in the mv-logic, we use the dual notion of good prefixes to define
the multi-valued safety property here. Of course, they are equivalent
in the classical case. We need $l$ to be complete in the following.

\begin{definition}\label{de:safety}

{\rm For an mv-linear-time property $P: (2^{AP})^{\omega}\rw l$,
define an mv-language $GPref(P): (2^{AP})^{\ast}\rw l$ as,

$GPref(P)(\theta)=\bigvee\{P(\theta\tau) | \tau\in
(2^{AP})^{\omega}\}$

\noindent for any $\theta\in (2^{AP})^{\ast}$. We call $GPref(P)$ the mv-language of
{\sl good prefixes} of $P$.

The {\sl closure} $Closure(P)$ of
$P$ is the mv-linear-time property over $(2^{AP})^{\omega}$ defined
as follows,

$Closure(P)(\sigma)=\bigwedge\{ GPref(P)(\theta)| \theta\in Pref(\sigma)\}$,

\noindent for any $\sigma\in (2^{AP})^{\omega}$, where
$Pref(\sigma)=\{\theta\in (2^{AP})^{\ast} |
\sigma=\theta\sigma^{\prime}$ for some $\sigma^{\prime}\in
(2^{AP})^{\omega}\}$ is called the prefix set of $\sigma$.

$P$ is called a {\sl safety property} if

$Closure(P)\subseteq P$.

\noindent }

\end{definition}

Informally, an mv-safety property can be characterized as ``anything always good must happen'', which is equivalent to the saying ``nothing bad should happen''.

An mv-safety property can be characterized by a closure operator which is
formally defined as follows.

\begin{proposition}\label{pro:closure}

For mv-linear-time properties $P, P_1$ and $P_2$, we have

(1)
$P\subseteq Closure(P)$;

(2) If $Im(P_1)$ and $Im(P_2)$ are finite
subsets of $l$, then $Closure(P_1\cup P_2)=Closure(P_1)\cup
Closure(P_2)$;

(3) $Closure(Closure(P))=Closure(P)$;

(4) $Closure(P)$ is the smallest safety property containing $P$, i.e., $Closure(P)$ is a safety property and if $Q$ is a safety property with $P\subseteq Q$, then $Closure(P)\subseteq Q$.

\end{proposition}

The proof is placed in Appendix B.

The following is immediately by Proposition \ref{pro:closure}(1) and the definition of safety property.

\begin{proposition}\label{pro:safety-closure}

For an mv-linear-time property $P$, $P$ is a safety property if and
only if $P=Closure(P)$.

\end{proposition}


Given $TS$, we define the finite trace function $Traces_{fin}(TS): (2^{AP})^{\ast}\rw l$ by letting $Traces_{fin}(TS)(\theta)=\bigvee\{Traces(TS)(\theta\tau) | \tau\in
(2^{AP})^{\omega}\}$ for any $\theta\in (2^{AP})^{\ast}$, i.e., $Traces_{fin}(TS)(\theta)=GPref(Traces(TS))(\theta)$. Then we obtain a useful implication of the mv-safety property as follows.

\begin{theorem}\label{th:safety}

Assume that $P$ is a safety property and $TS$ is an mv-TS. Then
$TS\models P$ if and only if $Traces_{fin}(TS)\subseteq GPref(P)$.

\end{theorem}

\noindent {\bf Proof:}\ \
``If'' part: Let $\sigma\in (2^{AP})^{\omega}$. We have
$Traces(TS)$ $(\sigma)\leq Traces_{fin}(TS)(\theta)$ for any $\theta\in
Pref(\sigma)$, and by assumption,
$Traces_{fin}(TS)(\theta)\leq GPref(P)(\theta)$. Hence, $Traces(TS)(\sigma)\leq \bigwedge \{GPref(P)(\theta) | \theta\in Pref(\sigma)\}=Closure(P)(\sigma)$, showing $Traces(TS)\subseteq Closure(P)$. Since $P$ is safe, $Closure(P)\subseteq P$ which implies $Traces(TS)\subseteq P$. Therefore, $TS\models P$.

``Only if'' part: Let $\theta\in (2^{AP})^*$. By assumption, for any $\tau\in (2^{AP})^{\omega}$, we have
$Traces(TS)(\theta\tau)\leq P(\theta\tau)$. So, $Traces_{fin}(TS)(\theta)=\bigvee\{Traces(TS)(\theta\tau) | \tau\in (2^{AP})^{\omega}\}\leq \bigvee\{P(\theta\tau) | \tau\in (2^{AP})^{\omega}\}=GPref(P)(\theta)$. Hence, $Traces_{fin}(TS)\subseteq GPref(P)$.\hfill$\Box$

Let us introduce an important mv-safety property, which is called
mv-invariance defined in the following manner.

\begin{definition}\label{de:invariant}

{\rm Let $\varphi$ be an mv-proposition formula generated by atomic
propositions in $AP$. A property $P: (2^{AP})^{\omega}\rw l$ is said
to be {\sl $\varphi$-invariant}, if
$P(A_0A_1A_2\cdots)=\bigwedge_{i\geq 0}\varphi(A_i)$ for any
$A_0A_1A_2\cdots\in (2^{AP})^{\omega}$.}

\end{definition}

For an mv-proposition formula $\varphi$ we let
$inv(\varphi): (2^{AP})^{\omega}\rw l$ be the property defined by $inv(\varphi)(A_0A_1A_2\cdots)=\bigwedge_{i\geq 0}\varphi(A_i)$ for any
$A_0A_1A_2\cdots\in (2^{AP})^{\omega}$.

\begin{proposition}\label{pro:invariant are safety}

Mv-invariance is an mv-safety property.

\end{proposition}

\noindent {\bf Proof:}\ \
If $P$ is $\varphi$-invariant, then $GPref(P):
(2^{AP})^{\ast}\rw l$ satisfies $GPref(P)$ $(A_0A_1\cdots
A_k)=\bigvee\{P(A_0A_1\cdots A_k\tau) | \tau\in
(2^{AP})^{\omega}\}\leq\bigwedge_{i=0}^k
\varphi(A_i)$.
Hence, $\bigwedge_{\theta\in
Pref(\sigma)}$\ \ $GPref(P)(\theta)\leq P(\sigma)$ for any $\sigma \in
(2^{AP})^{\omega}$. Therefore, $P$ is a safety
property. \hfill$\Box$

For an mv-proposition formula $\varphi$, and a finite mv-TS, $TS=(S,
Act, \rw, I, AP, L)$, we give an approach to reduce the model-checking problem
$TS\models inv(\varphi)$ into several classical model-checking problems of invariant properties.

For the given finite mv-TS, $TS=(S, Act, \rw, I, AP, L)$, let
$X=Im(I)\cup Im(\eta)$ and $l_1=<X>$, that is, $l_1$ is the
subalgebra of $l$ generated by $X$, then $l_1$ is finite as a set
(\cite{li93}). It is obvious that the behavior of $TS$ only takes
values in $l_1$. For this reason, we can assume that $l=l_1$
is a finite lattice in the following section. As just said in
Section 2, every element in $l$ can be represented as a join of
some join-irreducible elements of $l$.

For the given mv-transition system $TS=(S, Act, \rw, I, AP, L)$ and for
any $m\in JI(l)$, write $TS_m=(S, Act, \rw_m, I_m, AP, L)$, where
$\rw_m$ is the $m$-cut of $\rw$, i.e.,
$\rw_m=\{(s,\alpha,s^{\prime}) | \eta(s,\alpha, s^{\prime})\geq m\}$
and $I_m$ is the $m$-cut of $I$. Then $TS_m$ is a classical
transition system. By Proposition \ref{pro:resolution}, we have

$Traces(TS)=\bigcup_{m\in JI(l)}m\wedge Traces(TS_m)$.

For an mv-proposition formula $\varphi$ generated by the finite set $AP$, if we take $\varphi_m=\bigvee\{A\in 2^{AP}
|\varphi(A)\geq m\}$, then $\varphi_m$ is a classical proposition
formula. The classical safety property corresponding to $\varphi_m$,
denoted $inv(\varphi_m)$, is, $inv(\varphi_m)=\{A_0A_1\cdots |
\forall i. A_i\models \varphi_m\}=\{A_0A_1\cdots | \forall i.
\varphi(A_i)\geq m\}$. Noting that $inv(\varphi)_m=\{A_0A_1\cdots | \bigwedge_{i\geq
0}\varphi(A_i)\geq m\}=\{A_0A_1\cdots | \forall i. \varphi(A_i)\geq
m\}$, thus $inv(\varphi)_m=inv(\varphi_m)$. In this case, by
Proposition \ref{pro:resolution}, we have

$inv(\varphi)=\bigcup_{m\in JI(l)}m\wedge
inv(\varphi)_m=\bigcup_{m\in JI(l)}m\wedge inv(\varphi_m)$.

By Proposition \ref{pro:level of leq}, we have the following
observations:

$TS\models inv(\varphi)$ iff $Traces(TS)\subseteq inv(\varphi)$ iff
for all $m\in JI(l)$, $Traces(TS)_m\subseteq inv(\varphi)_m$, iff
for all $m\in JI(l)$, $TS_m\models inv(\varphi_m)$, iff for all
$m\in JI(l)$, $s\models \varphi_m$ for all states $s\in
Reach(TS_m)$, iff for all $m\in JI(l)$, $L(s)\models \varphi_m$ (in
proposition logic) for all states $s\in Reach(TS_m)$, where
$Reach(TS_m)$ denotes all the states reachable from the initial
states in $I_m$.

There are classical algorithms based on depth-first or width-first
graph search to realize $TS_m\models inv(\varphi_m)$ in
Ref.\cite{baier08}, and since $JI(l)$ is finite, then we can reduce
the mv-model-checking problem $TS\models \varphi$ into finite (in fact, at
most $|JI(l)|$) times of classical model-checking problems.

\begin{remark}\label{re:algorithm1}

{\rm The algorithm that implements the above reduction procedure
is placed in Algorithm 1. The classical model checker of invariant
properties is applied at most $|JI(l)|$ times. 
}

\end{remark}

\line(1,0){335}

{\bf Algorithm 1}: (Algorithm for the multi-valued model checking of
an invariant)

Input: An mv-transition system $TS$ and an mv-proposition formula
$\varphi$.

Output: return true if $TS\models inv(\varphi)$. Otherwise, return a
maximal element $x$ plus a counterexample for $\varphi_x$.

Set $A:=JI(l)$ \hfill(*The initial $A$ is the set of
join-irreducible elements of $l$*)

While ($A\not=\emptyset$) do

\ \ $x\longleftarrow$ the maximal element of $A$ \hfill(*$x$ is one
of the maximal elements of $A$*)

\ \  if  $TS_x\models inv(\varphi_x)$,   \hfill (*check if
$TS_x\models inv(\varphi_x)$ (using classical algorithm) is
satisfied *)

\ \  then


\ \ \ \ $A:=A-\{x\}$

\ \ else

\ \ \ \ Return $x$ plus a counterexample for $\varphi_x$   \hfill
(*if  $TS_x\not\models inv(\varphi_x)$, then there is a
counterexample for $\varphi_x$ *)

\ \ fi

od

 Return true

\line(1,0){335}

\subsection{Multi-valued liveness properties}

Compared to safety properties, ``liveness'' properties state that
¡±something good¡± will happen in the future. Whereas safety
properties are violated in finite time, i.e., by a finite system
run, liveness properties are violated in infinite time, i.e., by
infinite system runs. Related to multi-valued safety property, we
have multi-valued liveness property here.

\begin{definition}\label{de:liveness}

{\rm An mv-linear-time property $P: (2^{AP})^{\omega}\rw l$ is
called a {\sl liveness} property if $supp(Closure(P))=(2^{AP})^{\omega}$.}

\end{definition}

Similar to the classical liveness property, we have the following
proposition linking mv-safety and mv-liveness.

\begin{proposition}\label{le:safe-live}

For any mv-linear-time property $P: (2^{AP})^{\omega}\rw l$, there
exist an mv-safety property $P_{safe}$ and an mv-liveness property
$P_{live}$ such that $P=P_{safe}\cap P_{live}$.

\end{proposition}

\noindent {\bf Proof:}\ \ In fact, if we let $P_{safe}=Closure(P)$,
and $P_{live}=P\cup ((2^{AP})^{\omega}-supp(Closure(P)))$, then
$P=P_{safe}\cap P_{live}$ and
$supp(Closure(P_{live}))=(2^{AP})^{\omega}$. \hfill$\Box$




In the following, let us give some useful mv-liveness property used
in this paper.

\begin{definition}\label{de:persisitence}

{\rm Let $\varphi$ be an mv-proposition formula generated by atomical
proposition formulae $AP$, then the mv-{\sl persistence} property over
$AP$ with respect to $\varphi$ is the mv-linear time property $pers(\varphi):
(2^{AP})^{\omega}\rw l$ defined by,

$pers(\varphi)(A_0A_1\cdots)=\bigvee_{i\geq 0}\bigwedge_{j\geq i}\varphi(A_j)$.}

\end{definition}

Since we will use temporal modalities to characterize the
mv-persistence property, let us recall the semantics of two temporal
modalities $\lozenge$ (``eventually'', sometimes in the future) and
$\square$ (``always'', from now on forever) which are defined as
follows, for $A_0A_1\cdots\in (2^{AP})^{\omega}$, and a proposition
formula $\psi$ generated by atomic formulae $AP$,

$A_0A_1\cdots\models \lozenge\psi$ iff $\exists j\geq 0. A_j\models
\psi$;

$A_0A_1\cdots\models \square\psi$ iff $\forall j\geq 0. A_j\models
\psi$;

$A_0A_1\cdots\models \square\lozenge\psi$ iff $\forall i\geq
0.\exists j\geq i. A_j\models \psi$;

$A_0A_1\cdots\models \lozenge\square\psi$ iff $\exists i\geq
0.\forall j\geq i. A_j\models \psi$.

Now we give a characterization of the mv-persistence property $\varphi$ by its
cuts. Assume that $AP$ is finite. For $m\in Jl(l)$, as before, let $\varphi_m=\bigvee\{A\in 2^{AP} | \varphi(A)\geq m\}$. For the cut of $pers(\varphi)$, it is readily to verify that,
for any $m\in JI(l)$,

$pers(\varphi)_m=pers(\varphi_m)$,

\noindent where $pers(\varphi_m)$ is the classical persistence
property with respect to the proposition formula $\varphi_m$ generated by atomic
propositions $AP$, i.e.,

$pers(\varphi_m)=\{A_0A_1\cdots \in ((2^{AP})^{\omega} | \exists
i\geq 0.\forall j\geq i. A_j\models \varphi_m\}$.

Using the temporal operators, the above equality can be written as

$pers(\varphi_m)=\{\sigma\in ((2^{AP})^{\omega} | \sigma\models
\lozenge\square \varphi_m\}$.

By Proposition \ref{pro:resolution}, we have the following
resolution:

$pers(\varphi)=\bigcup_{m\in JI(l)} m\wedge pers(\varphi_m)$.

Then for an mv-TS, $TS$, by Proposition \ref{pro:level of leq}, we
have,

$TS\models pers(\varphi)$ iff $Traces(TS)\subseteq pers(\varphi)$
iff $\forall m\in JI(l)$, $Traces(TS)_m\subseteq
pers(\varphi)_m=pers(\varphi_m)$, iff $\forall m\in JI(l)$,
$TS_m\models pers(\varphi_m)$.

Then the mv-model checking $TS\models pers(\varphi)$ can be reduced
to at most $|JI(l)|$ times of  classical model checking $TS_m\models
pers(\varphi_m)$ for any $m\in JI(l)$. There is a nested
depth-first search algorithm to verify $TS_m\models pers(\varphi_m)$
(\cite{baier08}). Then the mv-model checking $TS\models
pers(\varphi)$ can be reduced to classical model checking.

We present the above reduction procedure in Algorithm 2. For
simplicity, we only write the different part of Algorithm 2
compared to Algorithm 1. Remark \ref{re:algorithm1} is also
applied to Algorithm 2.

\line(1,0){335}

{\bf Algorithm 2}: (Algorithm for the multi-valued model checking of
a persistence property)

Input: An mv-transition system $TS$ and an mv-proposition formula
$\varphi$.

Output: return true if $TS\models pers(\varphi)$. Otherwise, return
a maximal element $x$ plus a counterexample for $\varphi_x$.

Replace $TS_x\models inv(\varphi_x)$ by $TS_x\models
pers(\varphi_x)$ in the body of Algorithm 1.

 \line(1,0){335}

Mv-persistence property $pers(\varphi)$ is an mv-liveness property.
In fact, by Proposition \ref{pro:closure} (2),
$Closure(pers(\varphi))=Closure(\bigcup_{m\in JI(l)} m\wedge
pers(\varphi_m))=\bigcup_{m\in JI(l)} m\wedge
Closure(pers(\varphi_m))=\bigcup_{m\in JI(l)} m\wedge
(2^{AP})^{\omega}$, so $supp(Closure(pers(\varphi)))=(2^{AP})^{\omega}$.

The dual notion of mv-persistence property is called mv-dual
persistence property, which is defined as follows.

\begin{definition}\label{de:dual persistence}

{\rm Let $\varphi$ be an mv-proposition formula generated by atomical
proposition formulae $AP$, then the mv-{\sl dual persistence} property
over $AP$ with respect to $\varphi$ is the mv-linear time property
$dpers(\varphi): (2^{AP})^{\omega}\rw l$ defined by,

$dpers(\varphi)(A_0A_1\cdots)=\bigwedge_{i\geq 0}\bigvee_{j\geq i}\varphi(A_j)$.
}

\end{definition}

The duality of $pers$ and $dpers$ is shown in the following
proposition, which can be checked by a simple calculation.

\begin{proposition}\label{pro:dual}

$dpers(\varphi)=\neg pers(\neg \varphi)$.

\end{proposition}

Similarly to the property of $pers(\varphi)$, we have some
observations on the property of mv-dual persistence.

For the cuts of $dpers(\varphi)$, it is easy to verify that, for
any $m\in JI(l)$,

$dpers(\varphi)_m=dpers(\varphi_m)$,

\noindent where $dpers(\varphi_m)$ is the dual of the notion of
persistence property with respect to the proposition formula $\varphi_m$
generated by atomic propositions $AP$, i.e.,

$dpers(\varphi_m)=\{A_0A_1\cdots \in (2^{AP})^{\omega} | \forall
i\geq 0.\exists j\geq i. A_j\models \varphi_m\}$.

Then $dpers(\varphi_m)=\neg pers(\neg \varphi_m)$. Using the
temporal operators, we have

$dpers(\varphi_m)=\{\sigma\in (2^{AP})^{\omega} | \sigma\models
\square\lozenge \varphi_m\}$.

By Proposition \ref{pro:resolution}, it follows that

$dpers(\varphi)=\bigcup_{m\in JI(l)} m\wedge dpers(\varphi_m)$.

Then for an mv-TS, $TS$, by Proposition \ref{pro:level of leq}, we
have,

$TS\models dpers(\varphi)$ iff $Traces(TS)\subseteq dpers(\varphi)$
iff $\forall m\in JI(l)$, $Traces(TS)_m\subseteq
dpers(\varphi)_m=pers(\varphi_m)$, iff $\forall m\in JI(l)$,
$TS_m\models dpers(\varphi_m)$.

Then the mv-model checking $TS\models dpers(\varphi)$ can be reduced
to at most $|JI(l)|$ times of classical model checking $TS_m\models
dpers(\varphi_m)$ for any $m\in JI(l)$. As is well known, to check
$TS_m\models dpers(\varphi_m)$, it suffices to analyze the bottom
strongly connected components (BSCCs) in  $TS_m$ as a graph, which
will be done in linear time. That is to say, $A_0A_1\cdots\models
\square\lozenge B$ for a state subset $B\subseteq S$, iff $T\cap
B\not=\emptyset$ for each BSCC $T$ that is reachable from $s_0$,
where $L(s_0)=A_0$ and $s_0\in I_m$. For the detail, we refer to
Ref.\cite{baier08}.

We present the above reduction procedure in Algorithm 3. Remark
\ref{re:algorithm1} is also applied to Algorithm 3.

\line(1,0){335}

{\bf Algorithm 3}: (Algorithm for the multi-valued model checking of
a dual-persistence property)

Input: An mv-transition system $TS$ and an mv-proposition formula
$\varphi$.

Output: return true if $TS\models dpers(\varphi)$. Otherwise, return
a maximal element $x$ plus a counterexample for $\varphi_x$.

Replace $TS_x\models inv(\varphi_x)$ by $TS_x\models
dpers(\varphi_x)$ in the body of Algorithm 1.

\line(1,0){335}

\section{The verification of mv-regular safety property}
\label{sec:3}

In this and the next section, we shall give some methods of model checking of multi-valued
safety properties. We shall introduce an automata approach to check an mv-regular safety property by reducing it to checking
some invariant properties of a certain large system. In order to do this, let us first introduce the notion of finite
automaton in multi-valued logic systems, which are also called lattice-valued finite automaton in
this paper, please refer to Ref.\cite{li05,li07,li11} (c.f., Ref.\cite{droste12,demri07}).

\begin{definition}\label{def:l-vfa}
{\rm An {\sl $l$-valued finite automaton} ($l$-VFA for short) is a
5-tuple ${\cal A}=(Q,\Sigma,\delta,I,F)$, where $Q$ denotes a finite
set of states, $\Sigma$ a finite input alphabet, and $\delta$ an
$l$-valued subset of $Q\times \Sigma\times Q$, that is, a mapping
from $Q\times \Sigma\times Q$ into $l$, and $I$ and $F$ are $l$-valued subsets of $Q$,
that is, mappings from $Q$ into $l$, which represent the initial
state and final state, respectively. Then $\delta$ is called the
$l$-valued transition relation. Intuitively, $\delta$ is an
$l$-valued (ternary) predicate over $Q$, $\Sigma$ and $Q$, and for
any $p,q\in Q$ and $\sigma\in \Sigma$, $\delta(p,\sigma,q)$ stands
for the truth value of the proposition that input $\sigma$ causes
state $p$ to become $q$.  For each $q\in Q$, $I(q)$
indicates the truth value (in the underlying mv-logic) of the
proposition that $q$ is an initial state, $F(q)$ expresses the truth
value  of the proposition that $q$ is a final state.}

\end{definition}

The {\sl language} accepted by an $l$-VFA ${\cal A}$,  is the
mv-language $L({\cal A}): \sa\rw l$ defined as follows, for any
word $w=\sigma_1\sigma_2\cdots\sigma_k\in\sa$,

$L({\cal A})(w)=\bigvee\{I(q_0)\wedge\bigwedge_{i=0}^{k-1}
\delta(q_i,\sigma_{i+1},q_{i+1})\wedge F(q_{k})| q_i\in Q$ for any
$i\leq k\}$.

For an $l$-language $f: \sa\rw l$, if there exists an $l$-VFA ${\cal
A}$ such that $f=L({\cal A})$, then $f$ is called an {\sl $l$-valued regular language} or
{\sl mv-regular language} over $\Sigma$.

\begin{definition}\label{def:l-dfa}{\rm (c.f.\cite{li05})}
{\rm An {\sl $l$-valued deterministic finite automaton} ($l$-VDFA
for short) is a 5-tuple ${\cal A}=(Q,\Sigma,\delta,q_0,F)$, where
$Q$, $\Sigma$ and $F$ are the same as those in an $l$-valued finite
automaton, $q_0\in Q$ is the initial state, and the lattice-valued
transition relation $\delta$ is crisp and deterministic; that is,
$\delta$ is a mapping from $Q\times \Sigma$ into $Q$.}

\end{definition}

The {\sl language} accepted by an $l$-VDFA ${\cal A}$ has a simple form,
that is, for any word $w=\sigma_1\sigma_2\cdots\sigma_k\in\sa$, let
$q_{i+1}=\delta(q_i,\sigma_{i+1})$ for any $0\leq i\leq k-1$, then

$L({\cal A})(w)=F(q_k)$.

 Note that our definition of $l$-VDFA differs from the
usual definition of a deterministic finite automaton only in that
the final states form an $l$-valued subset of $Q$. This, however,
makes it possible to accept words to certain truth degrees (in the
underlying mv-logic), and thus to recognize mv-languages.

\begin{proposition}\label{pro:dfa}(\cite{li05,li07,li11})
$l$-VFA and $l$-VDFA are equivalent.

\end{proposition}

In fact, this result holds true for every bounded lattice $l$ (without any De Morgan and distributivity assumption), and even more general weight structures, c.f. \cite{CDIV,DV}.

We call an mv-safety property $P$ an {\sl mv-regular safety property}, if its mv-language of good prefixes $GPref(P)$ is
an mv-regular language over $2^{AP}$. For an mv-regular safety property $P$,
we assume that ${\cal A}$ is an $l$-VDFA accepting the good prefixes
of $P$, i.e., $L({\cal A})=GPref(P)$. This is a main difference
with the traditional setting of transition systems where
nondeterministic (finite-state or B\"{u}chi) automata do suffice.
The main reason is that we do not have the following implication
in multi-valued logic,

$A\leq B$ iff $A\wedge \neg B=\emptyset$.

\noindent So we need to verify $A\leq B$ directly instead of
checking $A\wedge \neg B=\emptyset$ as in classical case.

Now we give an approach to construct a new mv-TS from an mv-TS and
an $l$-VDFA.

\begin{definition}\label{de:product}

{\rm Let $TS=(S,Act,\rw,I,AP,L)$ be an mv-transition system without
terminal states and ${\cal A}=(Q,2^{AP},\delta,q_0,F)$ be an
$l$-VDFA with alphabet $2^{AP}$, the {\sl product transition system}
$TS\otimes {\cal A}$ is defined as follows:

$TS\otimes {\cal A}=(S^{\prime}, Act, \rw^{\prime}, I^{\prime},
AP^{\prime}, L^{\prime})$,

\noindent where $S^{\prime}=S\times Q$, $\rw^{\prime}$ contains all quadruples $((s,q),\alpha, (t,p), r)$ such that  $(s,\alpha, t, r)\in \rw$
(i.e., $\eta(s,\alpha, t)=r$) and $\delta(q, L(t))=p$;
$I^{\prime}(s_0,q)=I(s_0)$ if $\delta(q_0,L(s_0))=q$;
$AP^{\prime}=Q$ and $L^{\prime}: S^{\prime}\rw 2^{AP^{\prime}}$ is
given by $L^{\prime}(s,q)=\{q\}$.}

\end{definition}

Then for any $m\in JI(l)$, it can be readily verified that
$(TS\otimes {\cal A})_m=TS_m\otimes {\cal A}$.

Since ${\cal A}$ is deterministic, $TS\otimes {\cal A}$ can be
viewed as the unfolding of $TS$ where the automaton component $q$ of
the state $(s,q)$ in $TS\otimes {\cal A}$ records the current state
in ${\cal A}$ for the path fragment taken so far. More precisely,
for each (finite or infinite) path fragment $\pi=s_0s_1\cdots$ in
$TS$, there exists a unique run $q_0q_1\cdots$ in ${\cal A}$ for
$trace(\pi)=L(s_0)L(s_1)\cdots$ and
$\pi^{\prime}=(s_0,q_1)(s_1,q_2)\cdots$ is a path fragment in
$TS\otimes {\cal A}$. Vice verse, every path fragment in $TS\otimes
{\cal A}$ which starts in state $(s,\delta(q_0,L(s)))$ arises from
the combination of a path fragment in $TS$ and a corresponding run
in ${\cal A}$. Note that the $l$-VDFA ${\cal A}$ does not affect the
degree of trace function. That is, for each path $\pi^{\prime}$ in
$TS\otimes {\cal A}$ and its corresponding path $\pi$ in $TS$,
$Traces(TS\otimes {\cal
A})(trace(\pi^{\prime}))=Traces(TS)(trace(\pi))$. Then we have the
following theorem.

\begin{theorem}\label{th:verification regular safety}(The
verification of mv-regular safety property) For an mv-TS, TS, over
$AP$, let $P$ be an
mv-regular safety property over $AP$ such that $L({\cal
A})=GPref(P)$ for an $l$-VDFA ${\cal A}$ with alphabet $2^{AP}$. The following statements are equivalent:

(1) $TS\models P$;

(2) $Traces_{fin}(TS)\subseteq L({\cal A})$;

(3) $TS\otimes {\cal A}\models inv(\varphi)$, where
$\varphi=\bigvee_{q\in Q} F(q)\wedge q$.

\end{theorem}

\noindent {\bf Proof:}\ \ The equivalence of (1) and (2) has been
shown in Theorem \ref{th:safety}. To the end, it suffices to prove $(2)\Rightarrow (3)$ and
$(3)\Rightarrow (1)$.

For the $(3)\Rightarrow (1)$ part. Consider a path $\pi=s_0s_1s_2\cdots$ in $TS$ and any finite fragment
$\overline{\pi}=s_0\cdots s_n$ with
$\sigma=trace(\pi)=L(\pi)$ and
$\overline{\sigma}=trace(\overline{\pi})$. We claim that $Traces(TS)(\sigma)\leq
GPref(\overline{\sigma})=L({\cal A})(\overline{\sigma})$.
Then there is an infinite run $q_0q_1\cdots$ in ${\cal A}$ for
$\sigma$. Accordingly, $\delta(q_i,L(s_i))=q_{i+1}$ for
any $i\geq 0$. It follows that
$\pi^{\prime}=(s_0,q_1)(s_1,q_2)\cdots (s_n,q_{n+1})\cdots$ is an
infinite path in $TS\otimes {\cal A}$ with
$inv(\varphi)(L^{\prime}(\pi^{\prime}))=inv(\varphi)(\{q_1\}\{q_2\}\cdots)=\bigwedge_{i\geq
1} F(q_i)$. Then $Traces(TS)(\sigma)=Traces (TS\otimes {\cal
A})(L^{\prime}(\pi^{\prime}))\leq inv(\varphi)(L^{\prime}(\pi^{\prime}))=\bigwedge_{i\geq
1} F(q_i)$ by assumption. Hence, $Traces(TS)(\sigma)\leq F(q_{n+1})=L({\cal A})(\overline{\sigma})$ as claimed.

For the $(2)\Rightarrow (3)$ part. Consider any infinite run
$\pi^{\prime}=(s_0,q_1)(s_1,q_2)\cdots $. We claim that $TS\otimes {\cal
A}(L^{\prime}(\pi^{\prime}))\leq
inv(\varphi)(L^{\prime}(\pi^{\prime}))=\bigwedge_{i\geq 1}F(q_i)$.
Choose any $n$.
Then $\overline{\pi}=s_0\cdots s_n$ is a
finite fragment of $\pi=s_0s_1\cdots$ in $TS$ corresponding to
$\pi^{\prime}$. Furthermore, $\delta(q_i,L(s_i))=q_{i+1}$ for all
$i\geq 0$. It follows that $q_0\cdots q_{n+1}$ is an accepting run
for the $trace(s_0\cdots s_n)=L(s_0)\cdots L(s_n)=L(\overline{\pi})$
and $Traces(TS)(L(s_0)L(s_1)\cdots)$ $=Traces(TS\otimes {\cal
A})(L^{\prime}(s_0,q_1)L^{\prime}(s_1,q_2)\cdots)$. By assumption,
$Traces(TS)(L(\pi))\leq Traces_{fin}(TS)(L(\overline{\pi}))\leq L({\cal
A})(L(\overline{\pi}))=F(q_{n+1})$. Since $n$ was arbitrary, our claim follows. \hfill$\Box$

\begin{remark}\label{re:regular safety}

{\rm By Theorem \ref{th:verification regular safety}, for a regular
safety property $P$, to verify $TS\models P$, it suffices to check
$TS\otimes {\cal A}\models inv(\varphi)$, where ${\cal A}$ is an
$l$-VDFA satisfying $L({\cal A})=GPref(P)$, and $\varphi=\bigvee
F(q)\wedge q$. For the latter verification, we can use Algorithm 1 presented
in this paper.}

\end{remark}

\section{The verification of mv-${\omega}$-regular property}
\label{sec:4}

Now we further study some methods of model checking of multi-valued $\omega$-regular properties.
We need the notion of B\"{u}chi automata in multi-valued logic,
which can be found in Ref.\cite{kupferman07, DKR, DV}. We present this notion
with some minor changes.

\begin{definition}\label{de:buchi}

{\rm An {\sl $l$-B\"{u}chi automaton} ($l$-VBA, in short) is a 5-tuple ${\cal
A}=(Q,\Sigma,\delta,I,F)$ which is the same as an $l$-VFA, the
difference is the language accepted by ${\cal A}$, which is an
{\sl mv-$\omega$-language} $L_{\omega}({\cal A}): \Sigma^{\omega}\rw l$
defined as follows for any infinite sequence
$w=\sigma_1\sigma_2\cdots\in \Sigma^{\omega}$,

$L_{\omega}({\cal A})(w)=\bigvee\{I(q_0)\wedge\bigwedge_{i\geq
0}\delta(q_i,\sigma_{i+1}, q_{i+1})\wedge\bigwedge_{i\in J}F(q_j) |
q_i\in Q$ for any $i\geq 0$, and $J\subseteq {\bf N}$ is an infinite
subset of non-negative integers$\}$.

For an mv-$\omega$-language $f: \Sigma^{\omega}\rw l$, if there
exists an $l$-VBA ${\cal A}$ such that $f=L_{\omega}({\cal A})$,
then $f$ is called an {\sl mv-$\omega$-regular language} over $\Sigma$.

In an $l$-VBA ${\cal A}=(Q,\Sigma,\delta,I,F)$, if $\delta$ and $I$
are crisp, i.e., the image set of $\delta$ and $I$, denoted
$Im(\delta)$ and $Im(I)$ respectively, is a subset of $\{0,1\}$,
i.e., $Im(\delta)\subseteq \{0,1\}$ and $Im(I)\subseteq \{0,1\}$,
then ${\cal A}$ is called {\sl simple}. In this case, we also write
$Q_0=\{q\in Q | I(q)=1\}$ and $\delta(q,\sigma)=\{p\in Q |
\delta(q,\sigma,p)=1\}$.}

\end{definition}

If ${\cal A}$ is a simple $l$-VBA, then for any input
$w=\sigma_1\sigma_2\cdots\in \Sigma^{\omega}$, we have

$L_{\omega}({\cal A})(w)=\bigvee\{\bigwedge_{j\in J}F(q_j) | q_0\in
Q_0, q_j\in \delta(q_{j-1},\sigma_j)$ for any $j\geq 1$, and
$J\subseteq {\bf N}$ is an infinite
subset$\}=\bigvee\{\bigwedge_{i\geq 0}\bigvee_{j\geq i}F(q_j) |
q_0\in Q_0, q_j\in \delta(q_{j-1},\sigma_j)$ for any $j\geq 1\}$.

We shall show that each $l$-VBA is equivalent to a simple $l$-VBA in the
following.

Assume that ${\cal A}=(Q,\Sigma,\delta,I,F)$ is an $l$-VBA. Let
$X=Im(I)\cup Im(\delta)$, which is finite subset of $l$, and write
$l_1$ the sublattice of $l$ generated by $X$. Then $l_1$ is finite
as a set since $l$ is a distributive lattice. Construct a simple
$l$-VBA as, ${\cal
A}^{\prime}=(Q^{\prime},\Sigma,\delta^{\prime},Q_0^{\prime},F^{\prime})$,
where $Q^{\prime}=Q\times l_1$, and $\delta^{\prime}:
Q^{\prime}\times \Sigma\rw 2^{Q^{\prime}}$ is defined as,

$\delta^{\prime}((q,r),\sigma)=\{(p,s) | s=r\wedge
\delta(q,\sigma,p)\not=0$ for $p\in Q\}$;

\noindent $Q_0^{\prime}=\{(q,r) | r=I(q)\not=0\}$, and $F^{\prime}:
Q^{\prime}\rw l$ is, $F^{\prime}(q,r)=r\wedge F(q)$ for any
$(q,r)\in Q^{\prime}$.

For the new $l$-VBA, ${\cal A}^{\prime}$, for any input
$w=\sigma_1\sigma_2\cdots$,

$L_{\omega}({\cal A}^{\prime})(w)=\bigvee\{\bigwedge_{j\in
J}F^{\prime}(q_j,r_j) | (q_0,r_0)\in Q_0^{\prime}, (q_j,r_j)\in
\delta^{\prime}((q_{j-1},r_{j-1}),\sigma_j)$ for any $j\geq 1$, and
$J\subseteq {\bf N}$ is an infinite subset$\}$.

By a simple calculation, we can obtain that

$L_{\omega}({\cal A}^{\prime})(w)=\bigvee\{\bigwedge_{j\in
J}I(q_0)\wedge \delta(q_0,\sigma_1,
q_1)\wedge\cdots\wedge\delta(q_{j-1},\sigma_j,q_j)\wedge F(q_j) |
q_i\in Q$ for any $i\geq 0$ and $J\subseteq {\bf N}$ is an infinite
subset$\}=\bigvee\{I(q_0)\wedge\bigwedge_{i\geq
0}\delta(q_i,\sigma_{i+1}, q_{i+1})\wedge\bigwedge_{j\in J}F(q_j) |
q_i\in Q$ for any $i\geq 0$, and $J\subseteq {\bf N}$ is an infinite
subset of non-negative integers$\}=L_{\omega}({\cal A})(w)$.

Therefore, $L_{\omega}({\cal A})=L_{\omega}({\cal A}^{\prime})$,
${\cal A}$ and ${\cal A}^{\prime}$ are equivalent.

A simple $l$-VBA is called {\sl deterministic}, if $Q_0=\{q_0\}$ is a
single set and $\delta: Q\times \Sigma\rw Q$ is deterministic. As in
classical case, there is an $l$-VBA which is not equivalent to any
deterministic $l$-VBA.

In the case of deterministic $l$-VBA, the product of an mv-TS and a
deterministic $l$-VBA can also defined as before for the product of
mv-TS and an $l$-VDFA, the technique for mv-regular safety
properties can be roughly adopted.

\begin{theorem}\label{th:verification of infinite-regular property}
(The verification of mv-$\omega$-regular property using persistence)
Let $TS$ be an mv-TS without terminal states over $AP$ and let $P$
be an mv-$\omega$-regular property over $AP$ such that $L_{\omega}({\cal
A})=\neg P$ for a deterministic $l$-VBA ${\cal A}$ with the alphabet
$2^{AP}$. Then the following statements are equivalent:

(1) $TS\models P$;

(2) $TS\otimes {\cal A}\models pers(\varphi)$, where
$\varphi=\bigvee_{q\in Q} \neg F(q)\wedge q$.

\end{theorem}

\noindent {\bf Proof}\ \ For an infinite path $s_0s_1\cdots$ in
$TS$, since ${\cal A}$ is deterministic,
$q_{i+1}=\delta(q_i,L(s_i))$ is unique for any $i\geq 0$. Then it
follows that $P(L(s_0)L(s_1)\cdots)=\neg L_{\omega}({\cal
A})(L(s_0)L(s_1)$ $\cdots)=\neg(\bigwedge_{i\geq 0}\bigvee_{j\geq
i}F(q_j))=\bigvee_{i\geq 0}\bigwedge_{j\geq i} \neg F(q_j)$. On the
other hand, $pers(\varphi)(L(s_0,q_1)$
$L(s_1,q_2)\cdots)=pers(\varphi)(\{q_1\}\{q_2\}\cdots)=\bigvee_{i\geq
1}\bigwedge_{j\geq i}\neg F(q_j)$ $=\bigvee_{i\geq
0}\bigwedge_{j\geq i} \neg F(q_j)$. This shows that
$P=pers(\varphi)$. Noting that $Traces(TS)(L(s_0)$
$L(s_1)\cdots)=Traces(TS\otimes {\cal
A})(L(s_0,q_1)L(s_1,q_2)\cdots)$, it follows that
$Traces(TS)=Traces(TS\otimes {\cal A})$. Hence, condition (1) and
condition (2) are equivalent.  \hfill$\Box$

Dual to the above theorem, we can solve $TS\models P$ using an
mv-dual persistence property.

\begin{theorem}\label{th:dual-verification of infinite-regular property}

(The verification of mv-$\omega$-regular property using
dual-persistence) Let $TS$ be an mv-TS without terminal states over
$AP$ and let $P$ be an mv-$\omega$-regular property over $AP$ which can
be recognized by a deterministic $l$-VBA ${\cal A}$ with the
alphabet $2^{AP}$. Then the following statements are equivalent:

(1) $TS\models P$;

(2) $TS\otimes {\cal A}\models dpers(\varphi)$, where
$\varphi=\bigvee_{q\in Q} F(q)\wedge q$.

\end{theorem}

\begin{remark}\label{re:infinite-regular property}

{\rm Algorithm 2 and Algorithm 3 can be used for the verification
$TS\models P$ as presented in Theorem \ref{th:verification of
infinite-regular property} and Theorem \ref{th:dual-verification of
infinite-regular property}.}

\end{remark}

Since there are mv-$\omega$-regular properties which can not be
recognized by any deterministic $l$-VBA, Theorem \ref{th:dual-verification of infinite-regular property} does not
apply to the verification of all mv-$\omega$-regular properties. To
relax this restriction, we shall introduce another approach to the
verification of mv-$\omega$-regular properties. For this purpose, we
first introduce the notion of mv-deterministic Rabin automaton,
which is called  $l$-valued deterministic Rabin automaton here.

\begin{definition}\label{de:rabin}

{\rm An {\sl $l$-valued deterministic Rabin automaton} ($l$-VDRA, in
short) is a tuple ${\cal A}=(Q,\Sigma, \delta, q_0, {\cal F})$,
where $Q$ is a finite set of states, $\Sigma$ an alphabet, $\delta:
Q\times \Sigma\rw Q$ the transition function, $q_0\in Q$ the
starting state, and ${\cal F}: 2^Q\times 2^Q\rw l$.

A {\sl run} for $\sigma=A_0A_1\cdots\in \Sigma^{\omega}$ denotes an
infinite sequence $\rho=q_0q_1\cdots$ for states in ${\cal A}$ such
that $\delta(q_i,A_i)=q_{i+1}$ for $i\geq 0$. The run $\rho$ is
{\sl accepting} if there exists a pair $(H,K)\in 2^Q\times 2^Q$ such that
${\cal F}(H,K)>0$ and

$(\exists n\geq 0.\forall m\geq n. q_m\not\in H)\wedge (\forall
n\geq 0.\exists m\geq n. q_m\in K)$.

The {\sl accepted language} of ${\cal A}$ is a mapping $L_{\omega}({\cal
A}): \Sigma^{\omega}\rw l$, for any $\sigma=A_0A_1\cdots\in
\Sigma^{\omega}$,

$L_{\omega}({\cal A})(\sigma)=\bigvee\{{\cal F}(H,K) |$ there exists
an accepting run $\rho=q_0q_1\cdots$ such that $(\exists n\geq
0.\forall m\geq n. q_m\not\in H)\wedge (\forall n\geq 0.\exists
m\geq n. q_m\in K)\}$.}

\end{definition}

\begin{theorem}\label{th:vRA}

The class of mv-$\omega$-languages accepted by $l$-VDRAs is equal to
the class of mv-$\omega$-regular languages (those accepted by
$l$-VBAs).

\end{theorem}

We place the proof of this theorem at Appendix C.


Assume that $supp({\cal F})=\{(H_1,K_1), \cdots, (H_m,K_m)\}$ in the
following.

For an mv-transition system $TS=(S,Act,\rw,I,AP,L)$ and an mv-VDRA
${\cal A}=(Q,2^{AP},\delta,q_0,{\cal F})$, the product transition
system $TS\otimes {\cal A}$ is defined as follows:

$TS\otimes {\cal A}=(S^{\prime}, Act, \rw^{\prime}, I^{\prime},
AP^{\prime}, L^{\prime})$,

\noindent where $S^{\prime}=S\times Q$, $\rw^{\prime}$ contains all quadruples $((s,q),\alpha, (t,p), r)$ such that $(s,\alpha, t, r)\in \rw$
(i.e., $\eta(s,\alpha, t)=r$) and $\delta(q, L(t))=p$;
$I^{\prime}(s_0,q)=I(s_0)$ if $\delta(q_0,L(s_0))=q$;
$AP^{\prime}=2^Q$ and $L^{\prime}: S^{\prime}\rw 2^{AP^{\prime}}$ is
given by $L^{\prime}(s,q)=\{H\in AP^{\prime}=2^Q | q\in H\}$. In the
following, we write $\uparrow q=\{H\in AP^{\prime}=2^Q | q\in H\}$.

Let $Im({\cal F})-\{0\}=\{r_1,\cdots,r_m\}$ and ${\cal
F}_{[r_j]}=\{(H,K) | {\cal
F}(H,K)=r_j\}=\{(H_{j,1},K_{j,1}),\cdots$,
$(H_{j,m_j},K_{j,m_j})\}$. A related mv-(temporal-)proposition
formula about ${\cal A}$ is,

$\varphi=\bigvee_{j=1}^m r_j\wedge
\{\bigvee_{i=1}^{m_j}[(\lozenge\square\neg H_{j,i})\wedge
(\square\lozenge K_{j,i})]\}$.

The corresponding mv-linear-time property over $2^{AP^{\prime}}$ is
the mapping $d({\cal A}):(2^{AP^{\prime}})^{\omega}\rw l$, which is
defined as,

$d({\cal A})(A_0A_1\cdots)=\bigvee\{r_j | \exists i. (1\leq i\leq
m_j). (A_0A_1\cdots \models (\lozenge\square\neg H_{j,i})\wedge
(\square\lozenge K_{j,i}))\}=\bigvee\{r_j | \exists i. (1\leq i\leq
m_j). (\exists n\geq 0. \forall m\geq n. A_m\not\models
H_{j,i})\wedge (\forall n\geq 0.\exists m\geq n. A_m\models
K_{j,i})\}=\bigvee\{r_j | \exists i. (1\leq i\leq m_j). (\exists
n\geq 0. \forall m\geq n. H_{j,i}\not\in A_m)\wedge (\forall n\geq
0.\exists m\geq n. K_{j,i}\in A_m)\}$.

\begin{theorem}\label{th:verificaion infinite regular property}
(Verification of mv-$\omega$-regular property)

Let $TS$ be an mv-transition system over $AP$ without terminal
states, and let $P$ be an mv-$\omega$-regular property over $AP$
such that $L_{\omega}({\cal A})=P$ for some mv-VDRA ${\cal A}$. Then
the following statements are equivalent:

(1) $TS\models P$.

(2) $TS\otimes {\cal A}\models d({\cal A})$.

\end{theorem}

\noindent {\bf Proof}\ \ For a path
$\pi^{\prime}=(s_0,q_1)(s_1,q_2)\cdots$ in $TS\otimes {\cal A}$, its
projection to its first component $\pi=s_0s_1\cdots$ is a path in
$TS$. Since ${\cal A}$ is deterministic, the correspondence from
$\pi^{\prime}$ to $\pi$ is a one-to-one and onto mapping from the set
$Paths(TS\otimes {\cal A})$ to the set $Paths(TS)$. To complete the
proof, it suffices to show that the following two equations hold.

(i) $Traces(TS\otimes {\cal
A})(L^{\prime}(\pi^{\prime}))=Traces(TS)(L(\pi))$.

(ii) $d({\cal A})(L^{\prime}(\pi^{\prime}))=L_{\omega}({\cal
A})(L(\pi))$.

Let us prove the first equality. By the definition of $TS\otimes
{\cal A}$, we know

$Traces(TS\otimes {\cal
A})(L^{\prime}(\pi^{\prime}))=\bigvee\{\bigwedge_{i\geq 0} r_i |$
there exists $\alpha_1\alpha_2\cdots \in Act^{\omega}$,
$\pi_1=(s_0,q_1^{\prime})(s_1,q_2^{\prime})\cdots \in
(Q^{\prime})^{\omega}$, $r_0=I(s_0)$ and
$\eta^{\prime}((s_i,q_{i+1}^{\prime}),\alpha_{i+1}, (s_{i+1},
q_{i+2}^{\prime}))=r_{i+1}$ for any $i\geq 0$ and
$L^{\prime}(\pi^{\prime})=L^{\prime}(\pi_1)\}$.

Noting that $L^{\prime}(\pi^{\prime})=L^{\prime}(\pi_1)$ if and only
if $\uparrow q_i=\uparrow q_i^{\prime}$ for any $i$ and $\delta(q_0,
L(s_0))=q_1$. Since $\uparrow q_i=\uparrow q_i^{\prime}$ if and only
if $q_i=q_i^{\prime}$ by the definition of the operation $\uparrow$,
it follows that the run $\pi^{\prime}$ is uniquely defined by the
projected run $\pi=s_0s_1\cdots$. By the definition of $TS\otimes
{\cal A}$, we know $r_0=I(s_0)=I^{\prime}(s_0,q_1)$, and
$r_{i+1}=\eta^{\prime}((s_i,q_{i+1}),\alpha_{i+1},
(s_{i+1},q_{i+2}))=\eta(s_i,\alpha_{i+1}, s_{i+1})$. Hence,

$Traces(TS\otimes{\cal
A})(L^{\prime}(\pi^{\prime}))=\bigvee\{\bigwedge_{i\geq 0} r_i |$
there exists $\alpha_1\alpha_2\cdots \in Act^{\omega}$,
$\pi_1=s_0^{\prime}s_1^{\prime}\cdots \in S^{\omega}$,
$r_0=I(s_0^{\prime})$ and $\eta(s_i^{\prime}, \alpha_{i+1},
s_{i+1}^{\prime})=r_{i+1}$ for any $i\geq 0$ and
$L(\pi)=L(\pi_1)\}=Traces(TS)(L(\pi))$.

Therefore, $Traces(TS\otimes {\cal
A})(L^{\prime}(\pi^{\prime}))=Traces(TS)(L(\pi))$.

For the second equality, we know that

$d({\cal A})(L^{\prime}(\pi^{\prime}))=\bigvee\{ r_i |$ there exists
$i$, $1\leq i\leq m_j$, $L^{\prime}(\pi^{\prime})\models
\lozenge\square\neg H_{j,i}\wedge \square\lozenge
K_{j,i}\}=\bigvee\{{\cal F}(H_{j,i},K_{j,i}) |$
$L^{\prime}(\pi^{\prime})\models \lozenge\square\neg H_{j,i}\wedge
\square\lozenge K_{j,i}\}=\bigvee\{{\cal F}(H,K) |
L^{\prime}(\pi^{\prime})\models \lozenge\square\neg H\wedge
\square\lozenge K\}$.

We note that $L^{\prime}(\pi^{\prime})=\uparrow q_1\uparrow
q_2\cdots$ and $\delta(q_0,L(s_0))=q_1$. Then
\begin{eqnarray*}
&&L^{\prime}(\pi^{\prime})\models \lozenge\square\neg H\wedge
\square\lozenge K\\
 & {\rm if \ and \ only \ if}&
\uparrow q_1\uparrow q_2\cdots
\models \lozenge\square\neg H \ {\rm and}\ \uparrow q_1\uparrow q_2\cdots
\models \square\lozenge K\\
 & {\rm if \ and \ only \ if}&
(\exists n\geq 0.\forall
m\geq n. \uparrow q_m\models \neg H\ {\rm and}\ \forall n\geq 0.\exists
m\geq n. \uparrow q_m\models K) \\
& {\rm if \ and \ only \ if}&
(\exists n\geq
0.\forall m\geq n. q_m\not\in H \ {\rm  and}\ \forall n\geq 0.\exists m\geq
n. q_m\in K)\\
 & {\rm if \ and \ only \ if}&
{\rm the \ run}\ \rho=q_0q_1\cdots \ {\rm is \ an\
accepting\ run\ for\ the\ trace\ } \\
&& L(\pi)=L(s_0)L(s_1)\cdots.
\end{eqnarray*}

Hence, $d({\cal A})(L^{\prime}(\pi^{\prime}))=\bigvee\{{\cal F}(L,K)
| L^{\prime}(\pi^{\prime})\models \lozenge\square\neg H\wedge
\square\lozenge K\}=\bigvee\{{\cal F}(H,K) | (\exists n\geq
0.\forall m\geq n. q_m\not\in H) \wedge (\forall n\geq 0.\exists
m\geq n. q_m\in K)\wedge (\delta(q_0,L(s_0))=q_1\wedge
\delta(q_1,L(s_1))=q_2\wedge\cdots)\}=L_{\omega}({\cal
A})(L(s_0)L(s_1)\cdots)=L_{\omega}({\cal A})(L(\pi))$.

Therefore, $d({\cal A})(L^{\prime}(\pi^{\prime}))=L_{\omega}({\cal
A})(L(\pi))$. \hfill$\Box$

The verification of $TS\otimes {\cal A}\models d({\cal A})$ can also
be reduced to the classical model checking. Since $d({\cal
A})(L^{\prime}(\pi^{\prime}))=\bigvee\{{\cal F}(H,K) |
L^{\prime}(\pi^{\prime})\models \lozenge\square\neg H\wedge
\square\lozenge K\}$. It follows that $TS\otimes {\cal A}\models
d({\cal A})$ iff, for any $m\in JI(l)$, $(TS\otimes {\cal
A})_m\models \lozenge\square\neg H\wedge \square\lozenge K$ for
those $(H,K)$ such that $m\leq {\cal F}(H,K)$. Then the verification
of $TS\otimes {\cal A}\models d({\cal A})$ reduces to finite times
of classical model checking.

As is well known (\cite{baier08}), $(TS\otimes {\cal A})_m\models
\lozenge\square\neg H\wedge \square\lozenge K$ iff $(s,q_s)\models
\lozenge U$, where $q_s=\delta(q_0,L(s))$ for some $q_0\in I_m$, and
$U$ is the union of all accepting BSCCs in the graph of $(TS\otimes
{\cal A})_m$. A BSCC $T$ in $(TS\otimes {\cal A})_m$ is accepting if
it fulfills the acceptance condition ${\cal F}$. More precisely, $T$
is accepting iff there exists some $(H,K)\in {\cal F}_m$ such that

$T\cap (S\times H)=\emptyset$ and $T\cap (S\times K)\not=\emptyset$.

\noindent Stated in words, there is no state $(s,q)\in T$ such that
$q\in H$ and for some state $(t,q^{\prime})\in T$ it holds that
$q\in K$.

This result suggests determining the BSCCs in the product
transition system $(TS\otimes {\cal A})_m$ to check which BSCC is
accepting (i.e. determine $U$). This can be performed by a
standard graph analysis. To check whether a BSCC is accepting
amounts to checking all $(H,K)\in {\cal F}_m$. The overall complexity
of this procedure is

$O(|JI(l)|\times poly(size(TS), size({\cal A}))$£¬

\noindent where $size(TS)=|S|+|supp(\eta)|$, and $size({\cal
A})=|Q|+|supp(\delta)|$.

The related algorithm is presented in Algorithm 4. Remark
\ref{re:algorithm1} is also applied to Algorithm 4.

\line(1,0){335}

{\bf Algorithm 4}: (Algorithm for the multi-valued model checking of
an mv-$\omega$-regular property)

Input: An mv-transition system $TS$, an mv-$\omega$-regular
property $P$ and an $l$-VDRA ${\cal A}$ can accept $P$.

Output: return true if $TS\models P$. Otherwise, return a maximal
element $x$ plus a counterexample for $P_x$.

Set $A:=JI(l)$ \hfill(*The initial $A$ is the set of
join-irreducible elements of $l$*)

While ($A\not=\emptyset$) do

\ \ $x\longleftarrow$ the maximal element of $A$ \hfill(*$x$ is one
of the maximal element of $A$*)

\ \ ${\cal F}_x=\{(H,K) | {\cal F}((H,K))\geq x\}$ \hfill (*${\cal
F}_x$ is the $x$-cut of ${\cal F}$)

\ \  if  $(TS\otimes {\cal A})_x\models \bigwedge_{(H,K)\in {\cal
F}_x}\lozenge\square\neg H\wedge \square\lozenge K$,

\ \  then


\ \ \ \ $A:=A-\{x\}$

\ \ else

\ \ \ \ Return $x$ plus a counterexample for $(TS\otimes {\cal
A})_{x}\not\models \lozenge\square\neg H\wedge \square\lozenge K$
for some ${(H,K)\in {\cal F}_x}$   \hfill (*if $(TS\otimes {\cal
A})_x\not\models \bigwedge_{(H,K)\in {\cal F}_x}\lozenge\square\neg
H\wedge \square\lozenge K$, then there is a counterexample for
$(TS\otimes {\cal A})_{x}\not\models \lozenge\square\neg H\wedge
\square\lozenge K$ for some ${(H,K)\in {\cal F}_x}$*)

\ \ fi

od

 Return true

\line(1,0){335}

\section{Truth-valued degree of multi-valued model-checking}
\label{sec:5}

Another view and a more general picture of mv-model checking is focused on the membership degree
of mv-model checking as studied in Ref.\cite{chechik01}. Let us
recall its formal definition as follows.

\begin{definition}\label{de:membership model checking}

{\rm Let $P$ be an mv-linear-time property, and $TS$ an mv-TS. Then
the {\sl multi-valued model-checking function} is defined as,

$lMC(TS,P)= \bigcap _{\sigma\in
(2^{AP})^{\omega}}(\sigma\in Traces(TS)\rw \sigma\in P)$,

i.e.,

$lMC(TS,P)= \bigwedge\{Traces(TS)(\sigma)\rw P(\sigma) |\sigma\in
(2^{AP})^{\omega}\}$,

\noindent where $\rw$ is the implication operator in mv-logic.}

\end{definition}

{\bf Informally, the possibility of an mv-TS, $TS$, satisfying an mv-linear-time
property $P$, i.e., $lMC(TS$, $P)$, is the inclusion degree of
$Traces(TS)$ into $P$ as two mv-linear-time properties. In the definition of
$lMC(TS,P)$, the choice of the implication operator $\rw$ is in its
first importance. As remarked at the end of Section 2, there are two methods to determine the implication
operator. First, it can be defined by primitive connectives in
mv-logic system. For example, we can use $a\rw_m b=\neg a\vee b$ as a material implication or $a\rw_q b=\neg a\vee (a\wedge b)$ as a quantum logic implication to
define the implication operator. In fact, in
Ref.\cite{chechik04,chechik01}, the implication operator is chosen
as the material implication. They had some nice algebraic properties. However, this
definition can not grasp the essential of the function $lMC(TS,P)$
as indicating the inclusion degree of $Traces(TS)$ into $P$ as two
trace functions. In fact, intuitively, if $TS\models P$, we should
have $lMC(TS, P)=1$. But if we choose $a\rw_m b=\neg a\vee b$ or $a\rw_q b=\neg a\vee (a\wedge b)$, we would
not get $lMC(TS, P)=1$ even if $TS\models P$. For example, in
5-valued logic, $l$ is $l_5$ as shown in Fig. \ref{fig:mc1}, if we
choose $Traces(TS)\equiv U$ and $P\equiv L$, where $Traces(TS)\equiv U$ and $P\equiv L$
mean that $Traces(TS)(\sigma)= U$ and $P(\sigma)=L$ for any $\sigma\in
(2^{AP})^{\omega}$. Intuitively, we would get $TS\models P$, since $Traces(TS)(\sigma)=U<L=P(\sigma)$ for any $\sigma\in
(2^{AP})^{\omega}$, we would certainly get that if $\sigma$ satisfies $TS$, then $\sigma$ must also satisfy $P$. However, since
$U\rw_m L=\neg U\vee L=L\vee L=L$ and $U\rw_q L=\neg U\vee (U\wedge L)=L\vee U=L$, we would get $lMC(TS,P)=L$ but
not $lMC(TS,P)=1$. The verification result is too conservative  if we choose the implication operator as the material implication or the quantum logic implication. The second choice of the implication operator is
choosing $\rw$ as a primitive connective in mv-logic which satisfies
the condition $a\rw b=1$ whenever $a\leq b$ as we adopt in the paper. Back to the example just mentioned, since $TS\models P$, i.e., $Traces(TS)(\sigma)\leq P(\sigma)$ for any $\sigma\in
(2^{AP})^{\omega}$, it follows that $lMC(TS, P)=1$, just as we wanted. For more motivated examples, see the illustrative examples in next section.}

For the second choice of the implication operator, we need that $l$ is also a residual lattice. As said in Section 2, this is not a restriction. In fact, any finite De Morgan algebra is a residual lattice with implication operator defined as,

$a\rw b=\bigvee\{c | a\wedge c\leq b\}$.

For example, if $l$ is in linear order, then $a\rw b=1$ if $a\leq b$
and $a\rw b=b$ if $a>b$; if $l$ is a Boolean algebra, then $a\rw
b=\neg a\vee b$ as in the first case.

In particular, if $l=2$, then

$MC(TS,P)=lMC(TS,P)$.

The following proposition is simple, we present it here without
proof. Here, we choose the implication operator as the residual implication.

\begin{proposition}\label{pro:properties}

Let $TS$, $TS_1$ and $TS_2$ be mv-TS, $P$, $P_1$ and $P_2$ be mv-linear-time
properties. Then

(1) $lMC(TS,P)=1$ if and only if $TS\models P$.

(2)$lMC(TS,P_1\cap P_2)=lMC(TS,P_1)\cap lMC(TS,P_2)$.

(3)$lMC(TS,P_1)\vee lMC(TS,P_2)\leq lMC(TS,P_1\cup P_2)$.

(4) $lMC(TS_1+TS_2,P)=lMC(TS_1,P)\wedge lMC(TS_2,P)$, where $TS_1+TS_2$ is the disjoint union of $TS_1$ and $TS_2$. That is, for
$TS_i=(S_i,Act,\rw_i,I_i,L_i)(i=1,2)$, $TS_1+TS_2$ is
$(S,Act,\rw,I,L)$ with $S=S_1\times \{1\}\cup S_2\times
\{2\}$,

\begin{displaymath}
{\eta((s,i),\alpha,(t,j))}= \left\{ \begin{array}{ll}
\eta_i (s,\alpha,t), & \textrm{if $i=j$}\\
0 ,& \textrm{otherwise,}\\
\end{array} \right.
\end{displaymath}

\begin{displaymath}
{I_i((s,j))}= \left\{ \begin{array}{ll}
I_i(s), & \textrm{if $i=j$}\\
0 ,& \textrm{otherwise,}\\
\end{array} \right.
\end{displaymath}

\noindent and $L((s,i))=L_i(s) (i=1,2)$.

\end{proposition}

We give an approach to calculate $lMC(TS, P)$. Since $lMC(TS,
P)=\bigvee\{m\in JI(l) | m\leq lMC(TS,P)\}$, to calculate
$lMC(TS,P)$, it suffices to decide whether $lMC(TS, P)\geq m$ for
$m\in l$. Some analysis is presented as follows.

For $m\in l$, to decide $lMC(TS,P)\geq m$. Observe that

$lMC(TS,P)\geq m$

iff
$\bigwedge\{Traces(TS)(\sigma)\rw P(\sigma) | \sigma\in
(2^{AP})^{\omega}\}\geq m$,

iff
$\forall \sigma (2^{AP})^{\omega}$, $m\leq Traces(TS)(\sigma)\rw
P(\sigma)$,

iff
$\forall \sigma (2^{AP})^{\omega}$, $m\wedge Traces(TS)(\sigma)\leq
P(\sigma)$.

For $TS=(S, Act, \rw, I, AP, L)$ and $m\in L$, let $m\wedge TS=(S,
Act, \rw, I\wedge m, AP, L)$, where $I\wedge m: Q\rw l$ is defined
as, $I\wedge m(q)=I(q)\wedge m$ for any $q\in Q$. Then we have

$Traces(m\wedge TS)=Traces(TS)\wedge m$.

Hence, we have the following observation:

$\forall \sigma (2^{AP})^{\omega}$, $m\leq Traces(TS)(\sigma)\rw
P(\sigma)$

iff
$Traces(m\wedge TS)\subseteq P$,

iff
$m\wedge TS\models P$.

Thus, $lMC(TS,P)\geq m$ iff $m\wedge TS\models P$. We have presented algorithms
to decide $m\wedge TS\models P$ in Section 4 and Section 5. Hence it
is decidable whether $lMC(TS,P)\geq m$ holds for any $m\in JI(l)$.

The related algorithm for the calculation of $lMC(TS, P)$ is
presented as follows.

\line(1,0){335}

{\bf Algorithm 5}: (Algorithm for calculating $lMC(TS, P)$)

Input: An mv-transition system $TS$ and an mv-linear-time property
$P$.

Output: the value of $lMC(TS, P)$.

Set $A:=JI(l)$ \hfill(*The initial $A$ is the set of
join-irreducible elements of $l$*)

\ \ $B:=\emptyset$

While ($A\not=\emptyset$) do

\ \ $x\longleftarrow$ the maximal element of $A$ \hfill(*$x$ is one
of the maximal element of $A$*)

\ \  if  $x\wedge TS\models P$,   \hfill (*check if  $x\wedge
TS\models P$ (using Algorithm 1-4) is satisfied *)

\ \  then

\ \ \ \ $C:=\{y\in A | y\leq x\}$

\ \ \ \ $B:=B\cup C$

\ \ \ \ $A:=A-C$

\ \ else

\ \ \ \ $A:=A-\{x\}$

\ \ fi

od

 Return $``lMC(TS, P)="\ \bigvee B$

 \line(1,0){335}

\section{Illustrative examples and case study}
\label{sec:6}

Up to now, we have presented the theoretical part of model checking of linear-time properties in multi-valued logic. In this section, we give some examples to illustrate the methods of this article. First, we give an example to illustrate the constructions of this article. Then a case study is given.

\subsection{An example}

We now give an example to illustrate the construction of this
article. Note that this is a demonstrative rather than a case study
aimed at showing the scalability of our approach or the quality of
the engineering.

\begin{figure}[ptb]
\begin{center}
\includegraphics[width=\textwidth]{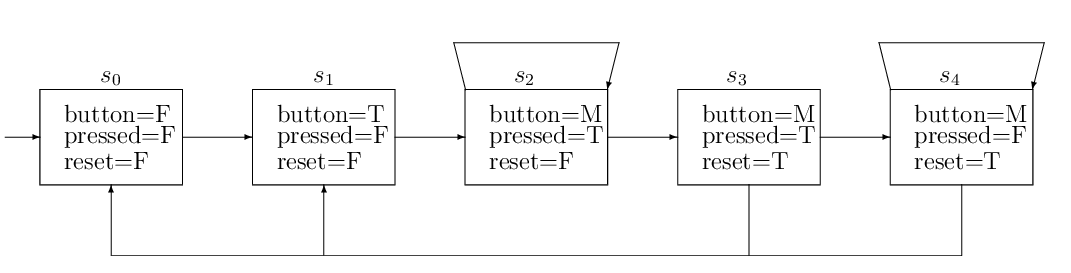}
\end{center}
\caption{State machine of the abstracted module Button in
Ref.\cite{chechik04}}
 \label{fig:mc2}
 \end{figure}

Consider the example of mv-transition system (in fact, mv-Kripke
structure, which can be considered as an mv-transition system with
only one internal action $\tau$) of the abstracted module Button
introduced in Ref.\cite{chechik04,chechik06} in 3-valued logic, which is
presented in Fig. \ref{fig:mc2}, where $l$ is the lattice $l_3$ of Fig.
\ref{fig:mc1}. This transition system has five states, $s_0, s_1,
s_2, s_3, s_4$, and the transition function is classical, i.e., with
values in the Boolean algebra $B_2=\{0,1\}$, here $0$=F, $1$=T. For
convenience, we only give those transitions with non-zero membership
values (as labels of the edge of the graph) in the following graph
representations of mv-transition systems and $l$-VDFA. For
simplicity, we only write those values of the labels of the edges
(corresponding to mv-transition) which are M. If there is no label of
the edges in the mv-transition system, then its value is T. The labeling
function of the mv-transition system is multi-valued, and there is only
one internal action $\tau$, the atomic propositions set is
$AP=\{$button, pressed, reset$\}$.


First, we transform this transition into its equivalent mv-TS with
ordinary labeling function as we have done in Appendix I, which is
presented in Fig. \ref{fig:mc3}. In Fig. \ref{fig:mc3}, $b, p$
and $r$ are short for the atomic propositions ``button'',
``pressed'', and ``reset'', respectively.

\begin{figure}[ptb]
\begin{center}
\includegraphics[width=\textwidth]{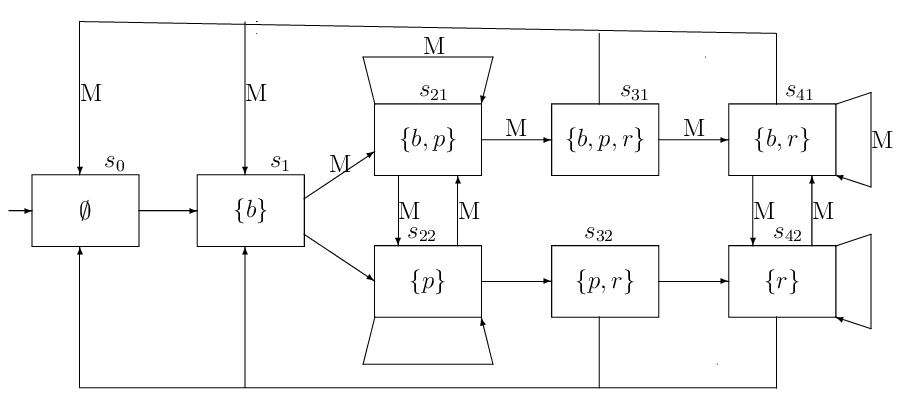}
\end{center}
\caption{Equivalent state machine $TS$ in Fig. \ref{fig:mc2} with
ordinary labeling function}
 \label{fig:mc3}
 \end{figure}


An mv-linear-time property $P: (2^{AP})^{\omega}\rw l$ is defined
by, for any $A_0A_1\cdots\in (2^{AP})^{\omega}$,

\begin{displaymath}
{P(A_0A_1\cdots)}= \left\{ \begin{array}{ll}
$T$, & \textrm{if $A_0=\emptyset$, $A_1=\{b \}$ and $A_i\not=\{b,p,r\}$ for any $i>1$}\\
$M$, & \textrm{if $A_0=\emptyset$, $A_1=\{b \}$ and $A_i=\{b,p,r\}$ for some $i>1$}\\
$F$ ,& \textrm{otherwise.}\\
\end{array} \right.
\end{displaymath}

Then the mv-language of good prefixes of $P$, $GPref(P): (2^{AP})^{\ast}\rw l$, is,

\begin{displaymath}
{GPref(P)(A_1\cdots A_{k})}= \left\{ \begin{array}{ll}
$T$ ,& \textrm{if $k=0$ or $k=1$ and $A_1=\emptyset$}\\
$T$,& \textrm{if $k\geq 2$, $A_1=\emptyset$, $A_2=\{b \}$ and $A_i\not=\{b,p,r\}$}\\
& \textrm {for any $i\leq k$}\\
$M$ ,& \textrm{if $k>2$ and $A_1=\emptyset$, $A_2=\{b \}$ and $A_i=\{b,p,r\}$}\\
& \textrm {for some $i\leq k$}\\
$F$ ,& \textrm{otherwise.}\\
\end{array} \right.
\end{displaymath}

It can be readily verified that $\bigwedge\{GPref(P)(\theta) |
\theta\in Pref(\sigma)\}=P(\sigma)$ for any $\sigma\in
(2^{AP})^{\omega}$, so $P$ is an mv-safety property.

\begin{figure}[ptb]
\begin{center}
\includegraphics[width=0.75\textwidth]{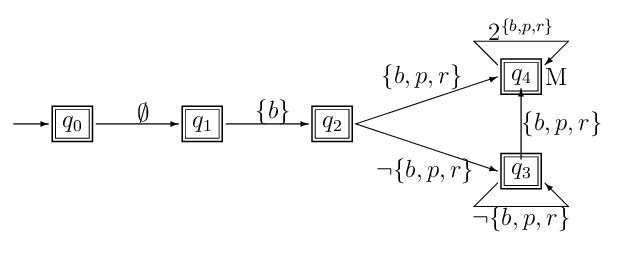}
\end{center}
\caption{An $l$-VDFA ${\cal A}$ which can recognize $GPref(P)$}
 \label{fig:mc4}
 \end{figure}

$GPref(P)$ is regular since it can be recognized by an $l$-VDFA
${\cal A}$ as presented in Fig. \ref{fig:mc4}. In ${\cal A}$, the
mv-final state $F$ is defined as, $F(q_0)=F(q_1)=F(q_2)=F(q_3)=$T,
and $F(q_4)=$M, as shown in Fig. \ref{fig:mc4}.

\begin{figure}[ptb]
\begin{center}
\includegraphics[width=\textwidth]{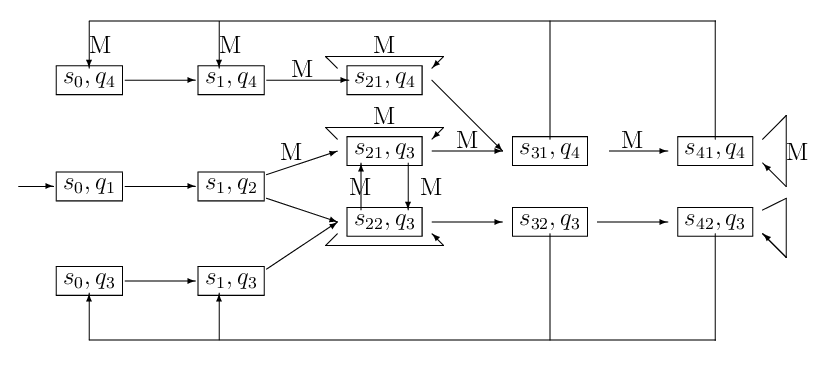}
\end{center}
\caption{The product transition system $TS\otimes {\cal A}$}
 \label{fig:mc5}
 \end{figure}

Then the product transition system $TS\otimes {\cal A}$ is presented
in Fig. \ref{fig:mc5}.

In the product transition system $TS\otimes {\cal A}$, the labeling
function is defined by $L^{\prime}(s,q)=\{q\}$ for any state
$(s,q)$, and $\varphi=q_1\vee q_2\vee q_3\vee$ M$q_4$. It can be
observed that $L^{\prime}(Reach((TS\otimes {\cal
A})_M))=\{q_1,q_2,q_3,q_4\}$, $L^{\prime}(Reach((TS\otimes {\cal
A})_{T}))=\{q_1,q_2,q_3\}$, $\varphi_M=q_1\vee q_2\vee q_3\vee q_4$
and $\varphi_{T}=q_1\vee q_2\vee q_3$. It is easily checked that,
for any $\alpha=$M or T, for any $(s,q)\in Reach((TS\otimes {\cal
A})_{\alpha})$, we have $L^{\prime}(s,q)=\{q\}\models \varphi_{\alpha}$. By
Theorem \ref{th:verification regular safety}, it follows that
$TS\otimes {\cal A}\models inv(\varphi)$ and thus $TS\models P$.

However, if we take $P^{\prime}=P\wedge$ M, that is,
$P^{\prime}(\sigma)=P(\sigma)\wedge$ M for any $\sigma\in
(2^{AP})^{\omega}$, $P^{\prime}$ is also an mv-safety property. If
we change $F$ in the above ${\cal A}$ into $F^{\prime}$, where
$F^{\prime}(q)=$M for any state $q$, and let the other parts
remain unchanged, then we obtain a new $l$-VDFA ${\cal A}^{\prime}$ such
that $L({\cal A}^{\prime})=GPref(P^{\prime})$. In this case, the
proposition formula $\varphi$ changes into
$\varphi^{\prime}=$M$q_0\vee$ M$q_1\vee$ M$q_2\vee$ M$q_3\vee$
M$q_4$ in $TS\otimes {\cal A}^{\prime}$. Then $TS_M\models
inv(\varphi^{\prime}_M)$ but $TS_{T}\not\models
inv(\varphi^{\prime}_{T})$. Since $\varphi^{\prime}_{T}=\bot$ and
$(s_1,q_3)\in Reach((TS\otimes {\cal A}^{\prime})_{T})$ but
$L^{\prime}(s_1,q_3)=\{q_3\}\not\models \bot=\varphi^{\prime}_{T}$,
which is a counterexample for the mv-model checking $TS\models
P^{\prime}$.

On the other hand, it is readily verified that M$\wedge TS\models
P^{\prime}$ but $TS\not\models P^{\prime}$. Hence $lMC(TS,
P^{\prime})$=M (by Algorithm \ref{fig:mc5}).

To apply Algorithm 4, we modify the $l$-VDFA in Fig.4 to make it an $l$-VDRA ${\cal B}$, where ${\cal F}:2^Q\times 2^Q\rw l$ is defined as, ${\cal F}(\emptyset, \{q_1,q_2,q_4\})=\top$, ${\cal F}(\{q_4\}, \{q_1,q_2,q_3\})=$M, and $\bot$ in other cases. Then ${\cal F}_{[\top]}=(\emptyset, \{q_1,q_2,q_4\})=\{(H_1,K_1)\}$, ${\cal F}_{[M]}=(\{q_4\}$, $\{q_1,q_2,q_3\})=\{(H_2,K_2)\}$. The corresponding mv-$\omega$-regular property $P''=L_{\omega}({\cal B})$ is defined as follows, for $\sigma=A_0A_1\cdots$,

\begin{displaymath}
{P''(\sigma)}= \left\{ \begin{array}{ll}
$T$ ,& \textrm{if $A_0=\emptyset$, $A_1=\{b\}$ and $A_2=\{b,p,r\}$}\\
$T$,& \textrm{if $A_0=\emptyset$, $A_1=\{b\}$,and there exists $k\geq 2$ such that $A_j\not=\{b,p,r\}$}\\
& \textrm{for $2\leq j\leq k$ and $A_{k+1}=\{b,p,r\}$ for any $i\leq k$}\\
$M$ ,& \textrm{if $A_0=\emptyset$, $A_1=\{b\}$ and $A_i=\{b,p,r\}$ for any $i\geq 2$}\\
$F$ ,& \textrm{otherwise.}\\
\end{array} \right.
\end{displaymath}

The structure of the product $TS\otimes {\cal B}$ is the same as the one in Fig. 5 except the labeling function.

Using Algorithm 4, it is easily checked that $(TS\otimes {\cal B})_{\top}\models\lozenge\square \neg H_1\wedge \square\lozenge K_1$ but $(TS\otimes {\cal B})_{M}\not\models\lozenge\square \neg H_2\wedge \square\lozenge K_2$, which is a counterexample for the model checking $TS\models P''$.

In fact, using Algorithm 5, we have $lMC(TS,P'')=$M.

\subsection{Case study}

In this section, we study how to verify a cache coherence protocol with the above methods.
Usually, in many distributed file systems, servers store files and clients store local copies of these files in their caches. Clients communicate with servers by exchanging messages and data (e.g., files) and clients do not communicate with each other. Moreover, each file is associated with exactly one authorized server.
There are two ways to ensure cache coherence. One is the client asks the server whether its copy is valid and the other is the server tells the client when the client's copy is no longer valid. Therefore, in a distributed system using a cache coherence protocol, if a client believes that a cached file is valid, then the server that is the authority on the file believes the client's copy is valid.

In this case study, we verify AFS2 (\cite{howard88}) that is a cache coherence protocol, which works as follows.

In the server, the initial state is $s_0$ at which the server believes the file is invalid. When the server receives the message $validate$ from the client and the file is valid, the server will transfer from $s_0$ to $s_1$ at which the server believes the file is valid, otherwise if the file is invalid, the server will still stay at $s_0$. Furthermore, the server will transfer from $s_0$ to $s_1$ when it receives the message $fetch$ from the client. In addition, the server will transfer from $s_1$ to $s_0$ when it receives the message $update$ from the client or the message $failure$, which respectively means that the client updates the file copy and the server needs to notify the other clients having the copy to update accordingly and there is something wrong in the communications between the client and server and they should check again the coherence of the file. It is represented in Fig.6.

\begin{figure}[ptb]
\begin{center}
\includegraphics[width=0.6\textwidth]{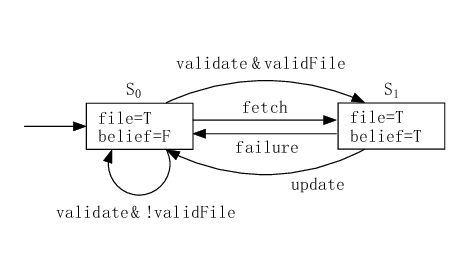}
\end{center}
\caption{The transition system of the server}
 \label{fig:mc6}
 \end{figure}

For the client, its initial states set are composed of $s_0$, $s_1$ and $s_2$. The state $s_0$ ($s_1$) represents that the client has no file copy in its cache and believes that the file is valid (invalid). The state $s_2$ describes that the client has a file copy and believes it is invalid. Therefore,
if the client starts as state $s_2$, it will send the message $val$ to ask the server whether or not the file copy in its cache is valid; while if the client starts as state $s_0$ or $s_1$, it will send the message $fetch$ to get the valid file directly from the server.
In addition, the state $s_3$ means that the client has a file copy and believes the file copy is valid. When the client receives the message $inval$ from the server, it will transfer from $s_3$ ($s_2$) to $s_0$ or $s_1$, which means that the server notifies the client that the copy is no longer valid and the client should discard the copy in its cache (as there is no file copy, so the validity of the file is unknown, i.e., the variable $belief$ equals either $true$ or $false$). When the client receives the message $failure$ from the system, it will transfer from $s_3$ to $s_2$, which means there is something wrong in the communications between the client and server and they should check again the coherence of the file. The transition system of a client is represented in Fig.7.

\begin{figure}[ptb]
\begin{center}
\includegraphics[width=0.5\textwidth]{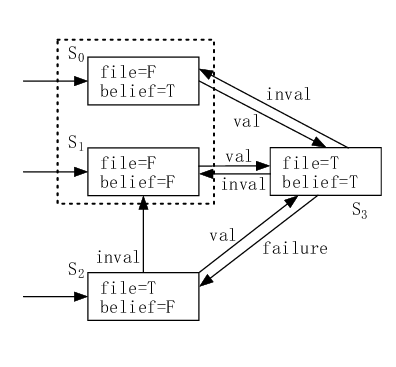}
\end{center}
\caption{The transition system of the client}
 \label{fig:mc7}
 \end{figure}

 In this case study, the pair of states $\{s_0,s_1\}$ of the client (indicated by dashed line in Fig.7) has a symmetric relation and this can be abstracted. This corresponds to the value of the variable $belief$ being irrelevant when the variable $file$ is $F$. Thus we can model the transition relation of the client by a 3-valued variable as shown in Fig.8.  When this model is composed with the rest of the AFS2 model, we get a 3-valued model-checking problem which can not be directly verified using a classical model-checking algorithm.

\begin{figure}[ptb]
\begin{center}
\includegraphics[width=0.6\textwidth]{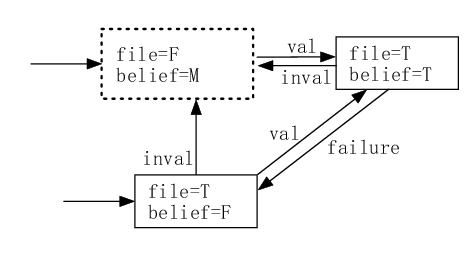}
\end{center}
\caption{The abstracted transition system of the client}
 \label{fig:mc8}
 \end{figure}

In addition, it might happen that the server sends an $inval$ message to some client that believes that its copy is valid. During the transmission, a property may hold since the client believes that its copy is valid while the server does not. Therefore, this transmission delay must be taken into account. We model the delay with the shared variable $time_i$.

The linear-time properties of AFS2 system we verified appeared as follows.

\line(1,0){335}

{\bf\sl P1:} If a client believes that a cached file is valid, then the server that is the authority on the file believes the client's copy is valid.

This property can be represented by a linear-temporal logic formulae as follows.

For one client:

$\square(Client_i.belief\wedge Client_i.file\rw (server.belief_i\wedge Server.file_i)\vee \neg time_i)\wedge (Server.out_i=val\rw Server.belief_i\wedge Sever.file_i)$.

For $N$ clients:

$\square(\bigwedge_{i=1}^N(Client_i.belief\wedge Client_i.file\rw (server.belief_i\wedge Server.file_i)\vee \neg time_i)\wedge (Server.out_i=val\rw Server.belief_i\wedge Sever.file_i))$.

{\bf\sl P2:} if a server believes that the client's copy is valid, then the client believes the cached file on the client is valid.

This property can be written as a linear-temporal logic formulae as follows.

For one client:

$\square(Server.belief_i\wedge Server.file_i\rw ((Client_i.belief\wedge Client_i.file)\vee \neg time_i)\wedge (Server.out_i=(validate\wedge valid-file)\vee fetch\rw Server.belief_i\wedge Sever.file_i)$.

For $N$ clients:

$\square(\bigwedge_{i=1}^N(Server.belief_i\wedge Server.file_i\rw ((Client_i.belief\wedge Client_i.file)\vee \neg time_i)\wedge (Server.out_i=(validate\wedge valid-file)\vee fetch\rw Server.belief_i\wedge Sever.file_i))$.

 \line(1,0){335}

The results are summarized in Fig.9, Table 1 and Table 2. The property $P1$ is correct, while the property $P2$ is wrong and a counterexample is given. There are several linear-temporal logic symbolic model checking tools as explained in Ref.\cite{rozier11}. The tool NuSMV 2.5.4 running on Pentium (R) Dual-Core E5800 with 3.20GHz processor and 2.00GB RAM, under ubuntu-11.04-desktop-i386, is used for the verification in this case study.

\begin{figure}[ptb]
\begin{center}
\includegraphics[width=0.7\textwidth]{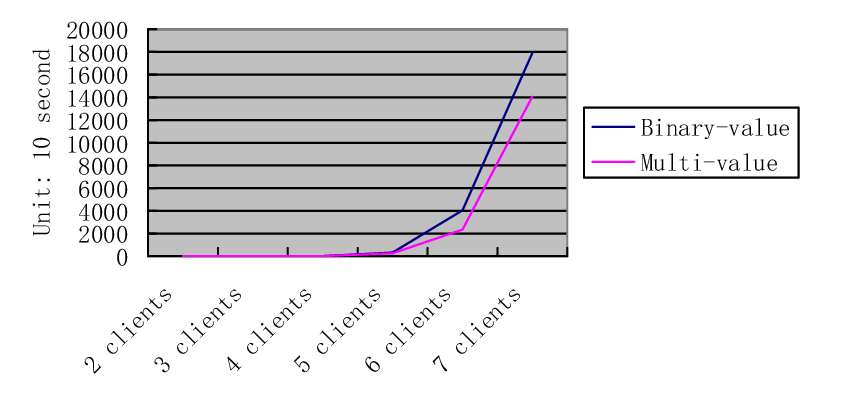}
\end{center}
\caption{The running times of the multi-valued and classical model checking for AFS2}
 \label{fig:mc9}
 \end{figure}

 \begin{table}
 \caption{The results of classical model checking for AFS2}
  \begin{centering}
\begin{tabular}{|l|l|l|l|l|}
  \hline
   & User Time & BDD Nodes & Transition Rules & States \\
    \hline
  2 Clients & 0.184s  & 33667 & $7^2$ & (2$\times 4)^2$ \\
      \hline
  3 Clients & 15.917s & 310383& $7^3$ & (2$\times4)^3$\\
      \hline
4 Clients & 20.845s & 1299115 & $7^4$& (2$\times4)^4$ \\
    \hline
 5 Clients & 322.224s & 235026 & $7^5$ & (2$\times 4)^5$ \\
     \hline
  6 Clients & 4054.901s & 443001 & $7^6$& (2$\times 4)^6$ \\
      \hline
  7 Clients &17885.806s & 1852283& $7^7$& (2$\times 4)^7$ \\
  \hline
\end{tabular}
\end{centering}
\end{table}

\begin{table}
\caption{The results of multi-valued model checking for AFS2}
  \begin{centering}
\begin{tabular}{|l|l|l|l|l|}
  \hline
   & User Time & BDD Nodes & Transition Rules & States \\
    \hline
  2 Clients & 0.1724s  & 33667 & $5^2$ & (2$\times 3)^2$ \\
      \hline
  3 Clients & 13.889s & 1061221& $5^3$ & (2$\times3)^3$\\
      \hline
4 Clients & 15.521s & 1360904 & $5^4$& (2$\times3)^4$ \\
    \hline
 5 Clients & 253.944s & 223831 & $5^5$ & (2$\times 3)^5$ \\
     \hline
  6 Clients & 2353.939s & 612687 & $5^6$& (2$\times 3)^6$ \\
      \hline
  7 Clients &14065.975s & 1318587& $5^7$& (2$\times 3)^7$ \\
  \hline
\end{tabular}
\end{centering}
\end{table}

In this case study, we use the classical model-checking algorithm two times to verify the model-checking problem of linear-time property in mv-logic. On the other hand, in classical model-checking of the original problem, the state space of the model is more complex than the abstracted model represented by mv-logic (as shown in Table 1 and Table 2). The overall time complexity of mv-logic is smaller than that in classical case as shown in Fig. 9, Table 1 and Table 2.

\section{Conclusions}
\label{sec:7}

Multi-valued model checking is a multi-valued extension to classical model checking. Both the model of the system and the
specification take values over a de Morgan algebra. Such an
extension enhances the expressive power of temporal logic and allows
reasoning under uncertainty. Some of the applications that can take
advantage of the multi-valued model checking are abstract
techniques, reasoning about conflicting viewpoints and temporal
logic query checking. In this paper, we studied several important
multi-valued linear-time properties and the multi-valued model
checking corresponding to them. Concretely, we introduced the
notions of safety, invariance, liveness,
persistence and dual-persistence in the multi-valued logic
system. Since the law of non-contradiction (i.e., $a\wedge \neg a=0$) and the law of excluded-middle (i.e., $a\vee \neg a=1$) do not hold in multi-valued logic, the linear-time properties introduced in this paper have new forms compared to those in classical logic. For example, the safety property in mv-logic is defined using good prefixes instead of bad prefixes. In which, model checking of the multi-valued invariant property
and the persistence property can be reduced to their classical
counterparts, the related algorithms were also presented.
Furthermore, we introduced the notions of lattice-valued finite automata including
B\"{u}chi and Rabin automata. With these notions, we gave the
verification methods of multi-valued regular safety properties and
multi-valued $\omega$-regular properties. Since the law of
non-contradiction and the law of excluded middle do not hold in
multi-valued logic, the verification methods gave here were direct
and not a direct extension of the classical methods. This was in contrast to the classical verification methods. A new form
of multi-valued model checking with membership degree (compared to
that in \cite{chechik01}) was also introduced. The related
verification algorithms were presented.

On the other hand, in literature there was much work on weighted model checking (\cite{buchholz03}, c.f.\cite{droste09}) that used weighted automata as models of systems. Weighted model checking uses a semiring as weight structure of weighted automata. Since a De-Morgan algebra is a distributive lattice, and a distributive lattice is a semiring, weighted model checking with weights in a De-Morgan algebra is a special case of semiring-weighted model checking. This kind of weighted model checking seems to be closed related with multi-valued model checking. However, they are different. There are some essential differences between multi-valued model checking and weighted model checking. First, weighted model checking is still based on classical logic, i.e., two-valued logic, while mv-model checking is based on  mv-logic. Then the uncertainty represented by the multi-valued logic systems can be considered sufficiently in multi-valued model checking. Second, there is few work on weighted LTL model checking, let alone the weighted model checking of the multi-valued safety
 property and liveness property, which formed the main topic of this paper. We should mention the recent paper \cite{li13}, in which the description of the classical linear-time properties using possibility measures was given, but not any work on the uncertainty linear-time properties, which was the topic of this paper.

There was much work on the multi-valued model checking, for example,
\cite{bruns99,bruns00,chechik00,chechik04,chechik012,chechik06,chechik01,godefroid03,huth01,kupferman07,droste09}.
As we said in the introduction part, we adopted a direct method to
model checking of multi-valued linear-time properties instead of
those existing indirect methods. More precisely, the existing methods of
mv-model checking still used the classical method with some minor
correction. That is, instead of checking $TS\models P$ for an
mv-linear time property $P$ using the inclusion of the trace
function $Traces(TS)\subseteq P$, the existing method only checked
the membership degree of the language $Traces(TS)\cap L({\cal
A}_{\neg P})$, where ${\cal A}_{\neg P}$ is an mv-B\"{u}chi
automaton such that $L({\cal A}_{\neg P})=\neg P$. However, as said
in Ref. \cite{baier08}, the equivalences and preorders between
transitions systems that ``correspond'' to linear temporal logic
are based on trace inclusion and equality. In this paper, we
adopted the multi-valued model checking of $TS\models P$ by using
directly the inclusion relation $Traces(TS)\subseteq P$. In general,
we used the implication connective as a primitive connective in mv-logic
which satisfies $a\leq b$ iff $a\rw b=1$ to define the membership
degree of the inclusion of $Traces(TS)$ into $P$. We give further comments on the comparison of our method to the existing approaches as follows.

Since we chose $\rw$ as a primitive connective in mv-logic, the classical logic could not be embedded into the mv-logic in a unique way as done in \cite{chechik06}. For example, $a\rw b$ and $\neg a\vee b$ are equivalent in  classical logic, but not in mv-logic. This is one of the main difference of our method to those existing approaches.
Due to this difference, we verify that the system model $TS$ satisfies the specified linear-time property $P$, i.e., $TS\models P$ directly using the inclusion $Traces(TS)\subseteq P$ instead of $L({\cal A})\cap L({\cal A}_{\neg P})=\emptyset$, where ${\cal A}_{\neg P}$ is a
multi-valued B\"{u}chi automaton such that $L({\cal A}_{\neg P})=\neg P$. Regarding expressiveness, we mainly studied the model-checking methods of linear-time properties in mv-logic systems. Compared with the work \cite{chechik01}, we use more general lattices instead of finite total order lattices to represent the truth values in the mv-logic. All the properties studied in  \cite{chechik01} can be tackled using our method, and another different view can be given. For the multi-valued model of CTL, etc, as done in \cite{chechik00,chechik012,chechik04,chechik06}, our method could be also applied which forms one direction of future work.

Therefore, the approach
proposed in this paper can be thought of as complementary to those
mentioned methods of multi-valued model checking. The examples and case study show the validity and performance of the method proposed in this article.
In the future work, we
shall give some further comparison of our method with those available
methods in multi-valued model checking and give some experiments. Another direction is to
extend the method used in this paper to multi-valued LTL or CTL.

\appendix
\section{The equivalent definition of
multi-valued transition system}

In an mv-TS, $TS=(S,Act, \rw, I, AP, L)$, if the labeling function
is $L: S\rw l^{AP}$ or $L: S\times AP\rw l$, then we have another
form of mv-TS. The later is used in Ref. \cite{chechik04} (which is
called mv-Kripke structure). There, $L(s,A)$ represents the
truth-value of the atomic proposition $A$ at state $s$.

In this case, the trace function of $TS$ needs to be redefined as
follows.

Since $TS$ is finite, we can assume that $Im(L)=\{d_1,\cdots,d_t\}$.
For any $d\in Im(L)$, define $L_d: S\rw 2^{AP}$ as follows,

$L_d(s)=\{ A\in AP | L(s,A)\geq d\}$.

Then $Traces(TS): (2^{AP})^{\omega}\rw l$ is defined in the
following manner. Let $A_0A_1\cdots\in (2^{AP})^{\omega}$,
$\rho=s_0\alpha_1 s_1\alpha_2\cdots$ a run of $TS$ with states
sequence $\pi=s_0s_1\cdots$, such that
$\eta(s_i,\alpha_{i+1},s_{i+1})=r_{i+1}$ and
$L_{d_{\phi(i)}}(s_i)=A_i$ for any $i\geq 0$, where $d_{\phi(i)}$ is
an element of $Im(L)$ with $\phi(i)\in \{1,\cdots,t\}$. Then,

$Traces(TS)(A_0A_1\cdots)=\bigvee\{r_0\wedge d_{\phi(0)}\wedge
r_1\wedge d_{\phi(1)}\wedge\cdots | \rho=s_0\alpha_1
s_1\alpha_2\cdots$ is a run of $TS$ with states sequence
$\pi=s_0s_1\cdots$, such that
$\eta(s_i,\alpha_{i+1},s_{i+1})=r_{i+1}$ and
$L_{d_{\phi(i)}}(s_i)=A_i$ for any $i\geq 0\}$.

We construct a new mv-TS from $TS$ with ordinary labeling function
which has the same traces function as the original mv-TS, $TS$.

Let $S^{\prime}=S\times\{1,\cdots,t\}$. The initial distribution
$I^{\prime}: S^{\prime}\rw l$ is defined by
$I^{\prime}(s,i)=I(s)\wedge d_i$, $\rw^{\prime}\subseteq
S^{\prime}\times Act\times S^{\prime}\times l$ is defined by
$\eta^{\prime}((s,i),\alpha,(s^{\prime},i^{\prime}))=d_i\wedge
\eta(s,\alpha,s^{\prime})\wedge d_{i^{\prime}}$, and $L^{\prime}:
S^{\prime}\rw 2^{AP}$ is defined by
$L^{\prime}(s,i)=L_{d_i}(s)=\{A\in AP | L(s,A)\geq d_i\}$. Then we
have a new mv-TS, $TS^{\prime}=(S^{\prime}, Act, \rw^{\prime},
I^{\prime}, AP, L^{\prime})$. Let us calculate the traces function
of $TS^{\prime}$ in the sequel.

For $A_0A_1\cdots\in (2^{AP})^{\omega}$,

$Traces(TS^{\prime})(A_0A_1\cdots)=\bigvee\{\bigwedge_{i\geq
0}r_i^{\prime} |$ there exists a run $\rho=s_0^{\prime}\alpha_1
s_1^{\prime}\alpha_2\cdots$ with states sequence
$\pi^{\prime}=s_0^{\prime}s_1^{\prime}\cdots$, such that
$\eta(s_i^{\prime},\alpha_{i+1},s_{i+1}^{\prime})=r_{i+1}^{\prime}$
and $L^{\prime}(s_i^{\prime})=A_i$ for any $i\geq 0\}$.

For a run $\rho=s_0^{\prime}\alpha_1 s_1^{\prime}\alpha_2\cdots$ in
$TS^{\prime}$, let $s_i^{\prime}=(s_i,\phi(i))$ and $d_{\phi(i)}\in
Im(L)$. Then from the definition of $I^{\prime}$, $\rw^{\prime}$,
and $L^{\prime}$, we know that

$r_0^{\prime}=I^{\prime}(s_0,\phi(0))=I(s_0)\wedge
d_{\phi(0)}=r_0\wedge d_{\phi(0)}$, where $r_0=I(s_0)$.

$r_i^{\prime}=\eta^{\prime}((s_{i-1},\phi(i-1)),\alpha_i,
(s_i,\phi(i)))=d_{\phi(i-1)}\wedge\eta(s_{i-1},\alpha_i,s_i)\wedge
d_{\phi(i)}=d_{\phi(i-1)}\wedge r_i\wedge d_{\phi(i)}$ for $i\geq
1$.

Thus, $\bigwedge_{i\geq 0}r_i^{\prime}=r_0\wedge d_{\phi(0)}\wedge
r_1\wedge d_{\phi(1)}\wedge\cdots$ and
$A_i=L^{\prime}(s_i^{\prime})=L_{\phi(i)}(s_i)$, which is the same
as those in the definition of $Traces(TS)(A_0A_1\cdots)$.

Hence, $Traces(TS^{\prime})(A_0A_1\cdots)=Traces(TS)(A_0A_1\cdots)$
for any $A_0A_1\cdots\in (2^{AP})^{\omega}$. It follows that
$Traces(TS^{\prime})=Traces(TS)$. Hence, $TS^{\prime}$ is equivalent to
$TS$ in the sense of trace function. \hfill$\Box$

\section{The proof of Proposition
\ref{pro:closure}}

(1) is obvious.

(2) The inclusion $Closure(P_1)\cup Closure(P_2)\subseteq
Closure(P_1\cup P_2)$ is obvious. Conversely, let $X=Im(P_1)\cup
Im(P_2)$, and let $l_1$ be the sublattice generated by $X$, then
$l_1$ is a finite distributive lattice (\cite{birkhoff40,li93}).
Observing that the three sets $Im(Closure(P_1))$, $Im(Closure(P_2))$ and
$Im(Closure(P_1\cup P_2))$ are subsets of $l_1$, to show
$Closure(P_1\cup P_2)\subseteq Closure(P_1)\cup Closure(P_2)$, it
suffices to show that, for any $m\in JI(l_1)$ and $\sigma\in
(2^{AP})^{\omega}$, $m\leq Closure(P_1\cup P_2)(\sigma)$ implies
that $m\leq Closure(P_1)(\sigma)$ or $m\leq Closure(P_2)(\sigma)$.
By the definition of $Closure$ operator, $m\leq Closure(P_1\cup
P_2)(\sigma)$ implies that, for any $\theta\in Pref(\sigma)$, there
exists $\tau\in (2^{AP})^{\omega}$ such that $m\leq
P_1(\theta\tau)\vee P_2(\theta\tau)$, it follows that $m\leq
P_1(\theta\tau)$ or $m\leq P_2(\theta\tau)$. Let $Pref_1=\{\theta\in
Pref(\sigma) | m\leq P_1(\theta\tau)$ for some $\tau\in
(2^{AP})^{\omega}\}$, and $Pref_2=\{\theta\in Pref(\sigma) | m\leq
P_2(\theta\tau)$ for some $\tau\in (2^{AP})^{\omega}\}$. Then
$Pref_1\cup Pref_2=Pref(\sigma)$. Since $Pref(\theta)$ is infinite
as a set, it follows that $Pref_1$ or $Pref_2$ is infinite. Without
loss of generality, let us assume that $Pref_1$ is infinite. Then,
for any $\theta\in Pref(\sigma)$, since $Pref_1$ is infinite, there
is $\theta_1\in Pref_1$ such that $\theta\in Pref(\theta_1)$, and
$m\leq P_1(\theta_1\tau_1)$ for some $\tau_1\in (2^{AP})^{\omega}$.
In this case, there exists $\tau\in (2^{AP})^{\omega}$ such that
$\theta_1\tau_1=\theta\tau$ and $m\leq
P_1(\theta_1\tau_1)=P_1(\theta\tau)$. Hence, by the definition of
$Closure(P_1)$, it follows that $m\leq Closure(P_1)(\sigma)$.

(3) By condition (1), we have $Closure(P)\subseteq
Closure(Closure(P))$. Conversely, for any $\sigma\in
(2^{AP})^{\omega}$, we have

$Closure(Closure(P))(\sigma)=\bigwedge\{\bigvee_{\tau\in
(2^{AP})^{\omega}}Closure(P)(\theta\tau) | \theta\in
Pref(\sigma)\}$.

On the other hand, for $Closure(P)(\theta\tau)$, since $\theta\in
Pref(\theta\tau)$, we have

$Closure(P)(\theta\tau)=\bigwedge\{\bigvee_{\tau_1\in
(2^{AP})^{\omega}}P(\theta_1\tau_1) | \theta_1\in Pref(\theta\tau)\}\leq
\bigvee_{\tau_1\in (2^{AP})^{\omega}}P(\theta\tau_1)$.

Hence,we have

$Closure(Closure(P))(\sigma)=\bigwedge\{\bigvee_{\tau\in
(2^{AP})^{\omega}}Closure(P)(\theta\tau) | \theta\in
Pref(\sigma)\}\leq \bigwedge\{\bigvee_{\tau\in (2^{AP})^{\omega}}$
$\bigvee_{\tau_1\in (2^{AP})^{\omega}}P(\theta\tau_1) | \theta\in
Pref(\sigma)\} =\bigwedge\{\bigvee_{\tau_1\in
(2^{AP})^{\omega}}P(\theta\tau_1) | \theta\in Pref$
$(\sigma)\}=Closure(P)(\sigma)$.

This shows that $Closure(Closure(P))\subseteq Closure(P)$.

Therefore, $Closure(Closure(P))= Closure(P)$. \hfill$\Box$

(4) is obvious.

\section{The proof of Theorem \ref{th:vRA}}

As a preliminary to show Theorem \ref{th:vRA}, we need a proposition
to characterize mv-$\omega$-regular languages. The following results are contained in \cite{DKR,DV}, we include a proof for its completeness.

\begin{proposition}\label{pro:infinite regular language}

For an mv-$\omega$ language $f:\Sigma^{\omega}\rw l$, the following
statements are equivalent:

(1) $f$ is an mv-$\omega$-regular language, i.e., $f$ can be
accepted by an $l$-VBA.

(2) $Im(f)$ is finite and $f_a$ is a $\omega$-regular language
(which can be accepted by a B\"{u}chi automaton) over $\Sigma$ for
any $a\in Im(f)$.

(3) There exist finite elements $m_1,\cdots, m_k$ in $l$ and finite
$\omega$-regular languages ${\cal L}_1,\cdots, {\cal L}_k$ over
$\Sigma$ such that

$f=\bigcup_{i=1}^km_i\wedge {\cal L}_i$.

\end{proposition}

\noindent {\bf Proof:}\ \ (1)$\Longrightarrow$ (2): Assume that $f$
is accepted by an $l$-VBA, ${\cal A}=(Q,\Sigma,\delta,I,F)$. Let
$X=Im(I)\cup Im(\delta)\cup Im(F)$. Since $Q$ and $\Sigma$ are
finite as two sets, $X$ is finite as a subset of $l$. Let $l_1$ be
the sublattice of $l$ generated by $X$, then $l_1$ is a finite
distributive lattice (\cite{birkhoff40,li93}), and any element of
$l_1$ can be represented as a finite join of join-irreducible
elements of $l_1$. For any $m\in JI(l_1)$, let ${\cal
A}_m=(Q,\Sigma,\delta_m,I_m,F_m)$. Then ${\cal A}_m$ is a classical
B\"{u}chi automaton and thus $L_{\omega}({\cal A}_m)$ is
$\omega$-regular.

Let us show that $L_{\omega}({\cal A})_m=L_{\omega}({\cal A}_m)$.
This is because, for any $w=\sigma_1\sigma_2\cdots\in
\Sigma^{\omega}$,

$w\in L_{\omega}({\cal A}_m)$

iff

for any $i\geq 0$, there exists $q_i\in Q$ such that $q_0\in I_m$,
$(q_i,\sigma_{i+1},q_{i+1})\in \delta_m$, and $J=\{i | q_i\in F_m\}$
is an infinite subset of {\bf N};

iff

 for any $i\geq 0$, there exists $q_i\in Q$ such that $I(q_0)\geq
m$, $\delta(q_i,\sigma_{i+1},q_{i+1})\geq m$, and there exists an
infinite subset $J$ of {\bf N} such that $F(q_j)\geq m$ for any
$j\in J$;

iff

for any $i\geq 0$, there exists $q_i\in Q$ and infinite subset $J$
of {\bf N} such that $I(q_0)\geq m$ and $\bigwedge_{i\geq
0}\delta(q_i,\sigma_{i+1},q_{i+1})\wedge \bigwedge_{j\in J} F(q_j)
\geq m$;

iff

$\bigvee\{I(q_0)\wedge \bigwedge_{i\geq
0}\delta(q_i,\sigma_{i+1},q_{i+1})\wedge \bigwedge_{j\in J} F(q_j) |
q_i\in Q$ for any $i\geq 0$ and $J$ is an infinite subset of {\bf
N}$\}\geq m$;

iff

$L_{\omega}({\cal A})(w)\geq m$

iff

$w\in L_{\omega}({\cal A})_m$.

Hence, $L_{\omega}({\cal A})_m$ is $\omega$-regular for any $m\in
JI(l_1)$.

Furthermore, for any $a\in Im(f)=Im(L_{\omega}({\cal A}))$, there
exists finite join-irreducible elements $m_1,\cdots,m_k$ in $l_1$
such that $a=\bigvee_{i=1}^k m_i$. Then

$f_a=\bigcap_{i=1}^k f_{m_i}$.

\noindent Since $f_{m_i}$ is $\omega$-regular and $\omega$-regular
languages are closed under finite intersection, it follows that
$f_a$ is $\omega$-regular.

$(2)\Longrightarrow$ (3) is obvious.

(3) $\Longrightarrow$ (1). Since ${\cal L}_i$ is $\omega$-regular,
there exists a B\"{u}chi automaton ${\cal
A}_i=(Q_i,\Sigma,\delta_i,I_i,F_i)$ such that $L_{\omega}({\cal
A}_i)={\cal L}_i$, for any $i=1,\cdots,k$. If we let $Q=\bigcup
\{i\}\times Q_i$, and define $I,F: Q\rw l$ and $\delta: Q\times
\Sigma\times Q\rw l$ as,

\begin{displaymath}
{I(i,q)}= \left\{ \begin{array}{ll}
m_i ,& \textrm{if $q\in I_i$}\\
0 ,& \textrm{otherwise,}\\
\end{array} \right.
\end{displaymath}

\begin{displaymath}
{F(i,q)}= \left\{ \begin{array}{ll}
m_i ,& \textrm{if $q\in F_i$}\\
0 ,& \textrm{otherwise,}\\
\end{array} \right.
\end{displaymath}

\begin{displaymath}
{\delta((i,q),\sigma,(j,p))}= \left\{ \begin{array}{ll}
m_i ,& \textrm{if $i=j$ and $(q,\sigma,p)\in\delta_i$}\\
0 ,& \textrm{otherwise,}\\
\end{array} \right.
\end{displaymath}

\noindent for any $(i,q), (j,p)\in Q$. This constructs a new
mv-$\omega$-B\"{u}chi automaton ${\cal A}=(Q,\Sigma,\delta,I,F)$.
Let us show that $L_{\omega}({\cal A})=f$.

In fact, for any $w=\sigma_1\sigma_2\cdots\in \Sigma^{\omega}$, for
any $i\geq 0$, if there exist $q_i^{\prime}\in Q$ and infinite
subset $J$ of {\bf N} such that $I(q_0^{\prime})\wedge
\bigwedge_{i\geq
0}\delta(q_i^{\prime},\sigma_{i+1},q_{i+1}^{\prime})\wedge
\bigwedge_{j\in J} F(q_j^{\prime}) > 0$. By definitions of $I,F$ and
$\delta$, there exists $j_i$, $1\leq j_i\leq k$ and $q_i\in Q$ such
that $q_i^{\prime}=(j_i,q_i)$ and $q_0\in I_{j_i}$,
$(q_i,\sigma_i,q_{i+1})\in \delta_{j_i}$, and for any $j\in J$,
$q_j\in F_{j_i}$. It follows that $w\in {\cal L}_{j_i}$. Hence, by
the definition of $L_{\omega}({\cal A})$, we have

$L_{\omega}({\cal A})(w)=\bigvee\{m_i | w\in {\cal L}_i\}=f(w)$.

Hence, $f$ is mv-$\omega$-regular. \hfill$\Box$

\begin{proposition}\label{pro:finite union}

Let $f_1, \cdots, f_k$ ($k\geq 2$) be finite mv-$\omega$-languages from
$\Sigma^{\omega}$ into $l$ which can be accepted by some $l$-VDRAs.
Then their join $f_1\cup\cdots\cup f_k$ can also be accepted by an
$l$-VDRA.

\end{proposition}

\noindent {\bf Proof:}\ \ For simplicity, we give the proof for the
case $k=2$. The other case can be proved by induction on $k$.

Assume that $f_i$ can be recognized by an $l$-VDRA ${\cal
A}_i=(Q_i,\Sigma, \delta_i,q_{i0},{\cal F}_i)$ for $i=1,2$,
respectively. Let us show that $f=f_1\cup f_2$ can also be accepted
by some $l$-VDRA. We explicitly construct such $l$-VDRA, ${\cal
A}=(Q,\Sigma,\delta,q_0,{\cal F})$, as follows, where $Q=Q_1\times
Q_2$, $\delta=\delta_1\times \delta_2$ (that is,
$\delta((q_1,q_2),\sigma)=(\delta(q_1,\sigma),\delta(q_2,\sigma))$),
$q_0=(q_{10},q_{20})$, and ${\cal F}: 2^{Q_1\times Q_2}\times
2^{Q_1\times Q_2}\rw l$ is defined by,

\begin{displaymath}
{{\cal F}((H,K))}= \left\{
\begin{array}{ll}
{\cal F}_1((H_1,K_1)) ,& \textrm{if $H=H_1\times Q_2$ and $K=K_1\times Q_2$}\\
{\cal F}_2((H_2,K_2)) ,& \textrm{if $H=Q_1\times H_2$ and $K=Q_1\times K_2$}\\
{\cal F}_1((H_1,K_1))\vee {\cal F}_2((H_2,K_2)) ,& \textrm{if $H=H_1\times Q_2\cup Q_1\times H_2$ and}\\
& \textrm{$K=K_1\times K_2$}\\
0 ,& \textrm{otherwise,}\\
\end{array} \right.
\end{displaymath}

By the definition of $L_{\omega}({\cal A})$, $L_{\omega}({\cal
A}_1)$ and $L_{\omega}({\cal A}_2)$, it is obvious that
$L_{\omega}({\cal A}_1)\cup L_{\omega}({\cal A}_2)\subseteq
L_{\omega}({\cal A})$.

Conversely, let $X=Im({\cal F}_1)\cup Im({\cal F}_1)$ and $l_1$ be
the sublattice generated by $X$, then $l_1$ is a finite distributive
lattice. The inclusion $Im({\cal F})\subseteq l_1$ is obvious and
thus $Im(L_{\omega}({\cal A}))\subseteq l_1$. To show
$L_{\omega}({\cal A})\subseteq L_{\omega}({\cal A}_1)\cup
L_{\omega}({\cal A}_2)$, it suffices to show that, for any
$\sigma\in \Sigma^{\omega}$ and for any $m\in JI(l_1)$, if $m\leq
L_{\omega}({\cal A})(\sigma)$, then $m\leq L_{\omega}({\cal
A}_1)(\sigma)$ or $m\leq L_{\omega}({\cal A}_2)(\sigma)$. By the
definition of $L_{\omega}({\cal A})(\sigma)$, if $m\leq
L_{\omega}({\cal A})(\sigma)$, then there exists $(H,K)\in 2^Q\times
2^Q$ such that $m\leq {\cal F}((H,K))$, and if we let
$q_{i+1}=(q_{1,i+1},q_{2,i+1})=\delta(q_i,\sigma)=(\delta_1(q_{1i},\sigma),\delta_2(q_{2i},\sigma))$
for $i=0,1,\cdots$, such that $(\exists n\geq 0.\forall m\geq
n.q_m\not\in H)\wedge (\forall n\geq 0.\exists m\geq n. q_m\in K)$.
By the definition of ${\cal F}$, we have three cases to consider:

Case 1: $H=H_1\times Q_2$, $K=K_1\times Q_2$. In this case, we have
$m\leq {\cal F}((H,K))={\cal F}((H_1,K_1))$. Then the sequence
$q_0q_1\cdots$ satisfies the condition $(\exists n\geq 0.\forall
m\geq n.q_m=(q_{1m},q_{2m})\not\in H_1\times Q_2)\wedge (\forall
n\geq 0.\exists m\geq n. q_m=(q_{1m},q_{2m})\in K_1\times Q_2)$. The
later condition implies that $(\exists n\geq 0.\forall m\geq
n.q_{1m}\not\in H_1)\wedge (\forall n\geq 0.\exists m\geq n.
q_{1m}\in K_1)$. By the definition of $L_{\omega}({\cal
A}_1)(\sigma)$, it follows that ${\cal F}_1((H_1,K_1))\leq
L_{\omega}({\cal A}_1)(\sigma)$. Hence, $m\leq L_{\omega}({\cal
A}_1)(\sigma)$.

Case 2: $H=Q_1\times H_2$, $K=Q_1\times K_2$. Similar to Case 1, we
can prove that $m\leq L_{\omega}({\cal A}_2)(\sigma)$.

Case 3: $H=H_1\times Q_2\cup Q_1\times H_2$ and $K=K_1\times K_2$.
In this case, we have $m\leq {\cal F}((H,K))={\cal F}((H_1,K_1))\vee
{\cal F}((H_2,K_2))$. Since $m\in JI(l_1)$, it follows that $m\leq
{\cal F}((H_1,K_1))$ or $m\leq {\cal F}((H_2,K_2))$. Consider the
sequence $q_0q_1\cdots$, it satisfies the condition $(\exists n\geq
0.\forall m\geq n.q_m=(q_{1m},q_{2m})\not\in H_1\times Q_2\cup
Q_1\times H_2)\wedge (\forall n\geq 0.\exists m\geq n.
q_m=(q_{1m},q_{2m})\in K_1\times K_2)$. The later condition implies
that $(\exists n\geq 0.\forall m\geq n.q_{1m}\not\in H_1)\wedge
(\forall n\geq 0.\exists m\geq n. q_{1m}\in K_1)\wedge (\exists
n\geq 0.\forall m\geq n.q_{1m}\not\in H_2)\wedge (\forall n\geq
0.\exists m\geq n. q_{1m}\in K_2)$. It follows that ${\cal
F}_1((H_1,K_1))\leq L_{\omega}({\cal A}_1)(\sigma)$ and ${\cal
F}_2((H_2,K_2))\leq L_{\omega}({\cal A}_2)(\sigma)$. Hence, $m\leq
L_{\omega}({\cal A}_1)(\sigma)$ or $m\leq L_{\omega}({\cal
A}_2)(\sigma)$.

This concludes that $L_{\omega}({\cal A})=L_{\omega}({\cal A}_1)\cup
L_{\omega}({\cal A}_2)$. \hfill$\Box$

{\bf The proof of Theorem \ref{th:vRA}:}

Let $f: \Sigma^{\omega}\rw l$ be an mv-language accepted by an
$l$-VDRA ${\cal A}=(Q,\Sigma,\delta,q_0,{\cal F})$. By the
definition of $L_{\omega}({\cal A})$, it follows that
$Im(f)=Im(L_{\omega}({\cal A}))\subseteq Im({\cal F})$ and thus
$Im(f)=Im(L_{\omega}({\cal A}))$ is a finite subset of $l$. For any
$a\in Im(f)$, $f_a$ is obvious accepted by the classical Rabin
automaton ${\cal A}_a=(Q,\Sigma,\delta,q_0,{\cal F}_a)$, and thus
$f=L_{\omega}({\cal A})$ is a $\omega$-regular language. Hence,
condition (2) in Proposition \ref{pro:infinite regular language}
holds for $f$, $f$ can be accepted by an $l$-VBA.

Conversely, if $f$ can be accepted by an $l$-VBA, then, by
Proposition \ref{pro:infinite  regular language}(3), there are
finite elements $m_1, \cdots, m_k$ in $l$ and finite
$\omega$-regular languages ${\cal L}_1,\cdots, {\cal L}_k$ over
$\Sigma$ such that

$f=\bigcup_{i=1}^k m_i\wedge {\cal L}_i$.

For any $i$, since ${\cal L}_i$ is $\omega$-regular, there exists a
deterministic Rabin automaton ${\cal A}=(Q, \Sigma,\delta,q_0$,
$ACC)$ accepting ${\cal L}_i$, i.e., $L_{\omega}({\cal A})={\cal
L}_i$. Construct an $l$-VDRA ${\cal A}^{\prime}$ from ${\cal A}$ as,
${\cal A}^{\prime}=(Q,\Sigma,\delta,q_0,{\cal F})$, where ${\cal F}:
2^Q\times 2^Q\rw l$ is,

\begin{displaymath}
{{\cal F}((H,K))}= \left\{
\begin{array}{ll}
m_i ,& \textrm{if $(H,K)\in ACC$}\\
0 ,& \textrm{otherwise.}\\
\end{array} \right.
\end{displaymath}

By a simple calculation, we have $L({\cal A}^{\prime})=m_i\wedge
{\cal L}_i$. This shows that $m_i\wedge {\cal L}_i$ can be accepted
by an $l$-VDRA for any $i$. By Proposition \ref{pro:finite union},
and the equality $f=\bigcup_{i=1}^k m_i\wedge {\cal L}_i$, it
follows that $f$ can be accepted by an $l$-VDRA. \hfill$\Box$



\end{document}